\documentclass[aps,pre,twocolumn,showpacs,floatfix,superscriptaddress]{revtex4}
\usepackage{graphicx}
\usepackage{amsmath}
\usepackage{latexsym}
\usepackage{epsfig}
\begin{document}

\title{Driven Disordered Polymorphic
             Solids: Phases and Phase Transitions, Dynamical
             Coexistence and Peak Effect Anomalies}

\author{Ankush Sengupta}
\affiliation{
Institut f\"ur Theoretische Physik II, Heinrich-Heine-Universt\"at,
Universit\"atsstra\ss{}e 1, 
D-40225 D\"usseldorf, Germany 
}
\author{Surajit Sengupta}
\affiliation{Centre for Advanced Materials, Indian Association for the Cultivation of Science,
2A \& 2B Raja S.C.  Mallik Road, Jadavpur, Kolkata, West Bengal 700 032, India
}
\affiliation{Advanced Materials Research Unit,
Satyendra Nath Bose National Centre for Basic Sciences, Block-JD, 
Sector-III, Salt Lake, Kolkata 700 098, India
}
\author{Gautam I. Menon}
\affiliation{
The Institute of Mathematical Sciences, CIT Campus, 
Taramani, Chennai 600 113, India
}
\date{\today}
\begin{abstract}
We study a simple model for the depinning and driven steady state phases
of a solid tuned across a polymorphic phase transition between ground
states of triangular and square symmetry. The competition between the underlying
structural phase transition in the pure system and the 
effects of the underlying disorder, as modified by the drive,
stabilizes a variety of unusual dynamical phases. These include
pinned states which may have
dominantly triangular or square correlations, a plastically flowing liquid-like phase,
a moving phase with hexatic correlations,
flowing triangular and square states and a dynamic
coexistence regime characterized by the complex interconversion of 
locally square and  triangular regions. We locate these phases in
a dynamical phase diagram and study them by defining and measuring appropriate
order-parameters and their correlations.  We demonstrate 
that the apparent power-law orientational correlations we obtain in our moving hexatic
phase arise from circularly averaging an orientational
correlation function
which exhibits long-range order in the (longitudinal) drive direction and short-range
order in the transverse direction. This calls previous simulation-based assignments of the
driven hexatic glass into question. The intermediate coexistence regime exhibits several
novel properties, including substantial enhancement in the current
noise, an unusual power-law spectrum of current fluctuations and striking metastability effects.
We show that this noise arises from the fluctuations of the interface
separating locally square and triangular ordered regions by demonstrating a
correlation between enhanced velocity fluctuations and local coordinations
intermediate between the square and triangular.  We demonstrate the
breakdown of effective ``shaking temperature''  treatments  in the coexistence regime by
showing that  such shaking temperatures are non-monotonic
functions of the drive in this regime. Finally we discuss the relevance of these simulations to the
anomalous behaviour seen in the peak effect regime of vortex lines in the disordered
mixed phase of type-II superconductors. We propose that  this anomalous behavior
is directly linked to the behavior exhibited in our simulations in the dynamical 
coexistence regime, thus suggesting a possible solution to the problem of the
origin of peak effect anomalies.
 \end{abstract} 
\maketitle

\section{Introduction}
The motion of an elastic medium across a quenched
disordered background presents a simple paradigm
for the understanding of several experiments\cite{fisherphysrep}. These include
studies of the depinning of charge-density waves\cite{gruner,thorne}, transport 
measurements in the mixed phase of type-II superconductors\cite{gia} as well as measurements
of the flow of colloidal particles across rough
substrates\cite{pertsinidis,coll-rand}. The issue of universality at
continuous depinning transitions has traditionally dominated
much of this literature, especially in the CDW context\cite{fisher85}. However, 
the behaviour of non-universal quantities in the vicinity of the 
depinning transition is often of  more
interest to the experimenter\cite{shobophysica}.  The nature of order,
correlations and response within the moving phase are also questions 
which underly many recent investigations of the physics of
non-equilibrium steady states, motivated in large part by the considerable
experimental literature on dynamical states of flux lines in
the mixed phase of driven, disordered type-II 
superconductors\cite{shobophysica}.

The canonical example of a remarkable {\em non-universal} feature of the depinning 
transition  is the peak effect often seen in the mixed phase in the
vicinity of the upper critical field $H_{c2}$\cite{belincourt}. The peak effect, 
a generic property of weakly disordered type-II superconductors, describes
the non-monotonic behaviour of the critical current $j_c$
as a function of the temperature $T$ or applied magnetic  field
$H$\cite{campbell}. This critical current measures the (depinning) force required
to induce observable motion of the vortex line array\cite{campbell,tinkham}. The peak effect is an often
spectacular phenomenon, with $j_c$ rising sharply in a narrow region whose width is
comparable to that of the zero-field superconducting transition\cite{shobophysica}.
Investigations of the peak effect describe a  host of unusual phenomena
associated with this narrow regime\cite{shobophysica}. 
These include ``finger print phenomena'', slow voltage oscillations, history dependent dynamic response,
enhanced low-frequency noise with a $1/f^\alpha$
spectrum and many other remarkable features. These are often
collectively referred to as ``peak effect anomalies''
\cite{shobo1, shobo2, shobo3, shobo4, shobo5, shobo6,ghosh, ravi1,satyajit1, satyajit2, satyajit4,andrei1,andrei2}.

Approaches to understanding the peak effect
have typically followed
two distinct paths. The first views the peak effect as
arising solely from the softening of the flux-lattice
close to H$_{c2}$, as reflected in the
vanishing of the shear elastic constant $C_{66} \sim (H
- H_{c2})^2$, while pinning strengths soften more gradually, as  $(H
- H_{c2})$\cite{pippard}. As suggested initially by Pippard\cite{pippard},  softer
lattices should be able to adapt better to random pinning\cite{larkin1,larkin2}. 
In the second class of theories, the peak effect is a reflection of 
the underlying phase diagram of a weakly
pinned, flux-line array in the $H-T$ 
plane\cite{gld0,gld,otterlo,nonomura,olsson,vinokur,chandan07,menon1,menon2,menon3,first04,thirdmom,cynthia2001a,cynthia2001b,brandt,baruchrmp}.
It has been argued that such phase diagrams should generically accomodate 
intermediate glassy  phases close to the melting transition\cite{vinokur,gld,satyajit4,menon1,menon2,menon3}. 
The peak effect  is then proposed to be associated with abrupt changes in
transport across such phase boundaries, with the anomalies rationalized  in terms of the glassy nature of such intermediate
states\cite{satyajit4,menon1,menon2,menon3,brandt}.  

A third, as yet unexplored, alternative to these approaches which addresses the
origin of the anomalies directly, combines the scenario  of an underlying static phase transition with the
possibility that driving such a system induces dynamical steady states with no static counterpart.
With this motivation in mind, the central questions addressed in this paper are 
the following:  Consider an underlying static phase transition in a pure system as
modified or broadened by weak quenched disorder. How are
signals of this transition manifest in dynamical
measurements? Further,  can novel dynamical
states with no analog in either the pure or the
disordered {\em undriven} system  be obtained once
the system is driven? Finally, could some of the remarkable
phenomenology of the peak effect  anomalies possibly originate in the properties
of such states?

We recently proposed a suitable model 
system capable of addressing some of these issues\cite{ankush, ankush1}. Our model 
uses interacting particles in two dimensions close to zero temperature. These
particles form a crystal in the absence of disorder. The
interaction potential contains a simple two-body
repulsive power-law  interaction as well as a short-range three-body
 interaction. The two-body interaction favours a triangular lattice. The three-body term
 favours a square lattice. We  tune between a square and a triangular ground
state by varying the strength of a single
parameter, the coefficient of the three-body interaction $v_3$.  We place
$N$ such interacting particles in a quenched disordered background, modeled numerically
in terms of a Gaussian random field with specified strength and two-point correlations. 
After finding the ground state of the interacting particles in the disordered background
using simulated annealing techniques, we apply a uniform driving force to the particles.
This results, first, in a depinning transition and then a sequence of partially ordered states
with varying degrees of spatial correlations as the force is increased.

The advantages to such a formulation are several. Signatures of phase transitions in
dynamical measurements are typically overwhelmed by  thermal fluctuations 
for purely temperature driven transitions, thereby obscuring the very effects we wish to characterize. 
Our model of a $T=0$ transition between square and triangular phases surmounts this problem
while also mirroring similar structural phase 
transitions in vortex lattices, typically between triangular and distorted rectangular phases, across which 
broadened peak effects are seen \cite{dewhurst,vinnikov,baruch1,white08,spie}. 

\begin{figure}
\begin{center}
\includegraphics[width=8cm]{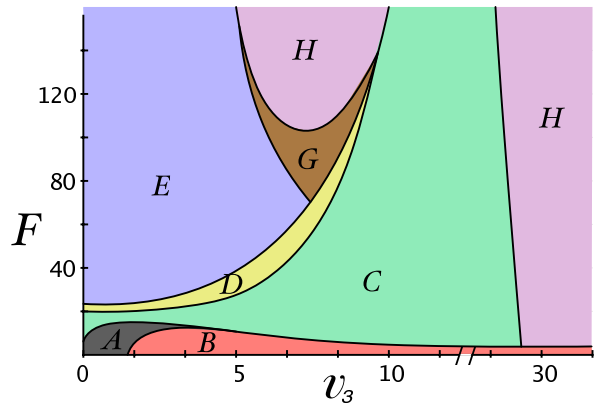}
\end{center}
\caption{
[Color Online]: Schematic Dynamical Phase Diagram of our model system, plotted as a function of
the three-body interaction strength $v_3$ and of driving force $F$ (see text for a more
detailed discussion). The phases are
(A) pinned triangular (B) pinned square (C) plastic flow (D) anisotropic 
hexatic (E) flowing triangular (G) coexistence regime  and (H) flowing square.
The precise location of the  lines separating these phases is disorder-dependent,
although the general topology of the phase diagram is not sensitive to disorder. The
boundary separating plastic flow from the flowing square phase is a very strong
function of the driving force $F$  at large $v_3$.
}
\label{phasedia}
\end{figure}

The sequence of steady states obtained in our model as a function of 
the uniform external driving force $F$  acting on the particles, and  for various values of the
three-body 
interaction strength  $v_3$, is summarized in the dynamical ``phase'' diagram
of Fig.~\ref{phasedia}. This phase diagram extends a similar phase
diagram proposed earlier to much larger
values of $v_3$\cite{ankush,ankush1}.  The phase diagram shows a variety
of phases: pinned states which may have dominantly
triangular or square correlations, a plastically flowing ``liquid-like'' state,
a moving anisotropic hexatic  phase, flowing
triangular and square states ordered over the size of our
simulation cell and a dynamic coexistence regime.

This paper presents a detailed characterization of these states. 
We  define and 
calculate appropriate correlation functions for $n-$atic (where $n=6$,
mostly) orientational order, in addition to structure factors measuring the distribution of
particles.  We capture local bond-orientational order in terms
of distribution functions of a complex number characterizing the local orientation.
This representation  is used to understand the action of 
an external force in biasing the
axes of orientational order. We demonstrate that the
hexatic order we measure arises as an artefact of averaging an
anisotropic quantity, which decays either exponentially to zero or to a
constant value  along two principal directions. In the ``ordered'' moving phases, square
and triangular, shown in  Fig.~\ref{phasedia}, orientational order  appears to be
established over length scales much large than our system sizes.

In the coexistence regime (labeled (G) in the phase diagram of Fig.~\ref{phasedia}),  
we investigate the nucleation and growth of ordered domains of one type (square or triangular) 
in another. Such nuclei are, in general, anisotropic,  forming along the
principal crystalline directions of each phase. The interface region of different domains is  
remarkably dynamic.  We assign  the substantial increase in noise we
see  within the coexistence regime
to the unusual properties of this interface, quantifying this proposal by linking
measures of fluctuation magnitude to coordinations intermediate between
square and triangular.

We measure several quantities on the dynamical side, including the
basic current-force relations, current statistics and
current noise at many different points in the
phase diagram. We calculate
the Koshelev-Vinokur ``shaking temperature''\cite{koshvin}, defined in detail
below, to understand how disorder-induced fluctuations in the flow
might provide an effective, pure-system temperature
in terms of which phase behaviour can be discussed. This shaking
temperature, while a monotonically decreasing function of the applied force
over much of the $F-v_3$ phase diagram, is strikingly {\em non-monotonic} in
$F$ within the coexistence regime. We correlate
local orientational and density fluctuations within the coexistence state  to
understand the origins of the anomalous noise in the coexistence
regime. We  compare the phenomenology 
of the simulations, specifically relating to the coexistence regime, 
to what is seen in experiments on peak effect anomalies in the mixed
state. We argue that this
comparison, supported by general phenomenological arguments, suggests
strongly that there may be a generic explanation for peak effect anomalies.

The outline of this paper is the following. In Section II, we describe the model
system we use, explaining our methodology and discussing relevant
features of the simulations and what we calculate.  In Section III, we provide
an overview of the phase diagram of the driven system, discussing, in particular detail,
the hexatic vortex glass
and the coexistence phase. We  discuss growth and
fluctuations of one phase within another, the behaviour upon
quenching and the origins of noise in this regime. In Section IV, we discuss
some aspects of peak effect anomalies seen in the experiments, pointing out
the close relationship between the diversity we see in our simulations with the experimental
data. We then conjecture that behaviour analogous to what
we see in the coexistence regime may be generic to all driven disordered
systems in the vicinity of an underlying static first-order phase transition in the pure limit.
Finally, in our concluding section, Section V, we summarize our results briefly and suggest
further lines of research.

\section{The Model System and Methodology}      
Our model system is two-dimensional and
consists of particles with two and three-body
interactions\cite{ankush,ankush1}. The three-body interaction,
parametrized through a single parameter $v_3$, tunes
the system across a square-triangular phase transition.
The total interaction energy for particles
confined to two-dimensions and labelled by their position 
vectors ${\bf r}_i$ is thus
\begin{equation}
V = 1/2\sum_{i\ne j}V_2(r_{ij})+
1/6\sum_{i \ne j \ne k} V_3(r_{i},r_{j},r_{k}),
\end{equation}
where $r_{ij} \equiv |{\bf r}_{ij}| \equiv |{\bf r}_j-{\bf r}_i|$.

We  take the two-body interaction to be of the power-law form
\begin{equation}
V_2(r_{ij})=v_{2}(\frac {\sigma_0}{r_{ij}})^{12},
\end{equation}
while the three-body term is
\begin{equation}
V_3(r_{i},r_{j},r_{k})=v_{3}[f_{ij}sin^{2}(4\theta_{ijk})f_{jk}+{\rm permutations}].
\end{equation}
The function 
$f_{ij} \equiv f(r_{ij}) = (r_{ij}- r_{0})^2$ for
$r_{ij}<1.8 \sigma_0$ and $0$ otherwise and $\theta_{ijk}$ is the angle
between ${\bf r}_{ji}$ and ${\bf r}_{jk}$.

The two-body interaction favors a triangular ground state while the three-body term favors 
$90^\circ$ and $45^\circ$ bonds and hence 
 a square structure.  Energy and length scales are set using $v_2 = 1$ and   
$\sigma_0 = 1$.  The zero-temperature phase diagram for particles
interacting with this potential  has been
calculated in Ref.~\cite{zerotemp}. As a function of
the parameter $v_3$, which measures the strength of
the three-body term, a discontinuous transition between a
triangular lattice, obtained for $v_3  < 1.5$, and a square
lattice, obtained for $v_3 > 1.5$, is seen.
A similar potential was  used
by Stillinger and Weber in an early study of
melting of a square solid\cite{stillinger}.

Particles also interact with background quenched disorder in the
form of  a one-body Gaussian random potential field
V$_d({\bf r})$ with zero
mean and exponentially decaying (short-range) correlations. This potential field
is defined on a fine grid following a methodology due to Chudnovsky and 
Dickman\cite{chudnovsky}, and interpolation is used to find the value of the potential at 
intermediate points\cite{epl}. The disorder variance is
set to $v_d^2 = 1$ and its spatial correlation length is $\xi=0.12$. Larkin length
estimates\cite{larkin1,larkin2,gia} yield $L_a/a  \sim 100$, with $a =
1/{\rho}^{1/2}$ the lattice parameter, somewhat larger
than our system size.

\subsection{Methodology}
The system evolves through standard Langevin dynamics;
\begin{eqnarray}
\dot{{\bf r}_i} &=& {\bf v}_i, \nonumber \\
\dot{{\bf v}_i} &=& {{\bf f}_i}^{\rm int} - \gamma {\bf v}_i + {\bf F} + {\bf
\eta}_i(t).
\end{eqnarray}
Here ${\bf v}_i$ is the velocity, ${{\bf
f}_i}^{\rm int}$ the total interaction force, and ${\bf
\eta}_i(t)$ the random force acting on particle $i$,
simulating thermal fluctuations at temperature $T$. A
constant force ${\bf F} = \{F_x,0\}$ 
drives the system. The zero-mean thermal noise ${\bf \eta}_i(t)$
is specified by 
\begin{equation}
<{\bf \eta}_i(t){\bf }{\bf \eta}_j(t')> = 2T\gamma\delta_{ij}\delta(t-t'),
\end{equation}
with  $T=0.1$, well below the equilibrium melting temperature of
the system. The unit of time $\tau = 
\gamma \sigma_0^2/v_2$, with $\gamma = 1$ the viscosity. 

\subsection{Simulation details}
Our system consists of $N$ particles, with $N$ between $1600$ and   $10,000$, in a
square box at number density $\rho = 1.1$.  
Configurations obtained through a simulated annealing
procedure are the initial inputs to our Langevin simulations. 
This annealing procedure involves equilibration using a NVT Monte
Carlo scheme at fixed density and within a fixed background
potential of a tunable amplitude. The strength of the disorder
is then increased in steps to the working disorder strength, with the
system equilibrated at each step for around $10^5$ Monte Carlo
steps. Varying the strength of disorder in this fashion enables the
system to converge to its true minimum energy state more
efficiently than methods which employ a temperature annealing
schedule.

We evolve the system using a time step of $10^{-4}\tau$.
The external force $F_x$ is ramped up from a
starting value of 0, with the
system maintained at upto $10^8$ steps at 
each $F_x$.  

Given the local instantaneous particle density
\begin{equation}
\rho({\bf r},t) = \sum_i \delta({\bf r} - r_i(t)), 
\end{equation}
we calculate a variety of structural observables
at equal time, such as the static
structure factor $S({\bf q})$ defined by
\begin{equation}S({\bf q}) = \sum_{ij} \exp(-i{\bf q \cdot r}_{ij}).
\end{equation}
where ${\bf r}_{ij} =  {\bf r}_i - {\bf r}_j$.
Delaunay triangulations yield
the probability distributions $P(n)$ 
of $n=4,5,6$ and $7$ coordinated 
particles ($\sum_n P(n) = 1$)\cite{preparata}.  We define 
order parameters
\begin{equation}
\psi = (P(4)-P(6))/(P(4) + P(6)),
\end{equation}
to distinguish between square 
and triangular phases,
\begin{equation}
\psi_{\Delta} = (P(6)-P(5)-P(7))/(P(6)+P(5)+P(7)),
\end{equation}
to 
distinguish between liquid (disordered) and triangular crystals and 
\begin{equation}
\psi_{\Box} = (P(4)-P(5)-P(7))/(P(4)+P(5)+P(7)),
\end{equation}
to distinguish between 
liquid and square crystals. In addition, we compute the  
hexatic order parameter 
\begin{equation}
\psi_{6,i} \equiv \psi_{6}({\bf r}_{i}) = \sum_{j} \exp(-i6\theta_{ij}),
\end{equation}
and its correlations, defined via
\begin{equation}
g_6(r) = \langle \psi_6(0)\psi_6(r)\rangle
\end{equation}
where $\theta$ defines the bond angle 
associated with the vector connecting neighbouring particles,
as measured with respect to an arbitrary external axis. The second
moment of the  distribution of $\psi_{6,i}$
is  the bond-orientational susceptibility. Below, we
describe alternative distribution functions which quantify the
extent to which the axes of the crystal align along the driving
force direction. These include ``Argand plots'' of the distribution of
a complex  quantity which characterizes local orientational
order. 
\begin{figure*}
\begin{center}
\includegraphics[width=18.5cm]{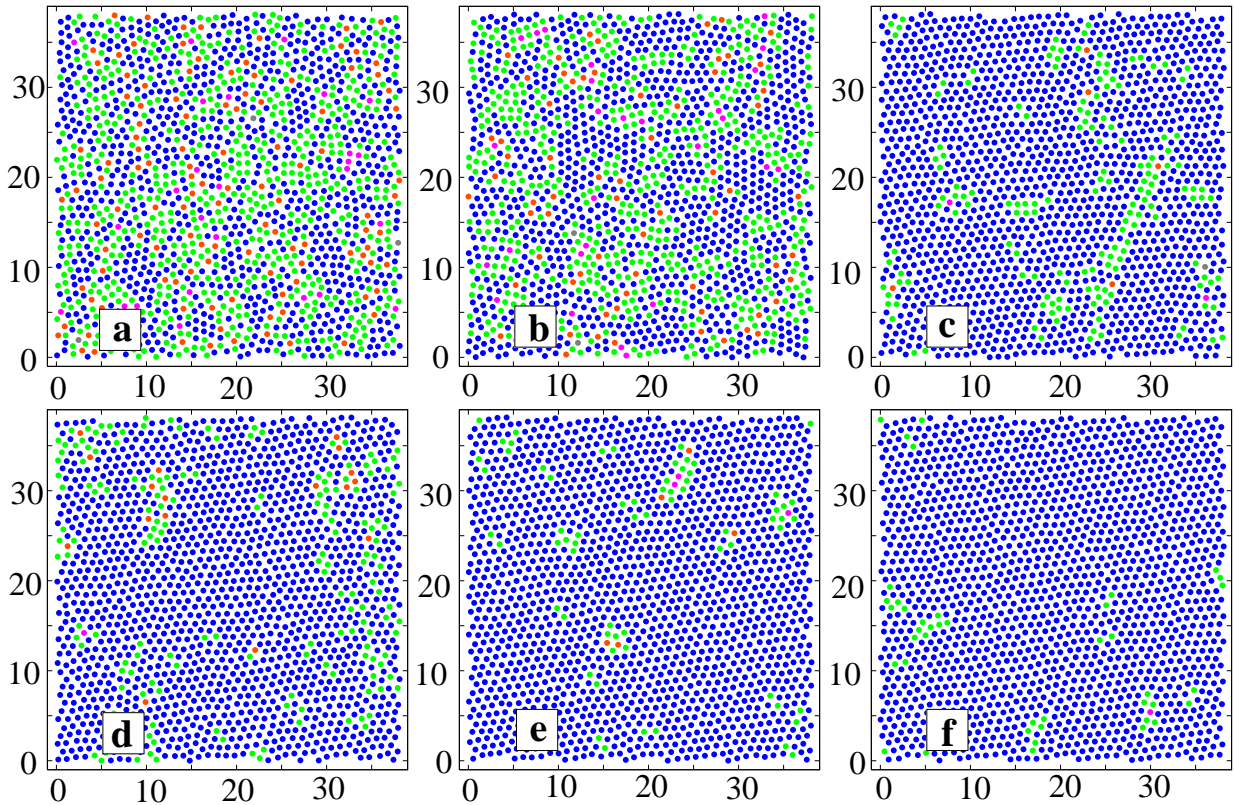}
\end{center}
\caption{
[Color Online]: Instantaneous snapshots of steady state configurations of the system 
at  $T=0.1$ with $v_3 = 6.0$. 
The configurations (a) - (f)  
are respectively for driving forces $F_x = 6, 10, 20, 22, 32$ and $40$ respectively. 
The particles are colored according to their 
coordination number $n$ = 4 (magenta), 5 (green), 6 (blue), 7 (orange)
and 8 (grey),  as obtained from Delaunay triangulations. 
}
\label{Recryst}
\end{figure*}
The dynamical variables we study include the center
of mass velocity $v_{cm}$,  and 
the particle flux and its statistics.
The centre-of-mass velocity is defined via
\begin{equation}
v_{cm} =  \langle  \frac{1}{N} \sum_i v_i(t) \rangle,
\end{equation}
where the brackets $\langle \cdot \rangle$ denote an average
in steady state and $v_i(t)$ is the velocity of particle $i$ at time $t$. 
We measure the particle flux by counting the
number of particles which cross an imaginary line crossing the
$x=0$ axis in a single time step and then averaging this
result over time. This flux, essentially a current 
$j(x,y)$ integrated over all $y$ at fixed x, which must be
independent of $x$ in steady state, has fluctuations about a constant
value. We measure  and discuss the power spectrum of these fluctuations.

Koshelev and 
Vinokur (KV)\cite{koshvin} have suggested that the combination of the
drive and the disorder should  yield an effective
``shaking'' temperature in the moving phase. Such an effective
temperature manifests itself in 
transverse and longitudinal fluctuations of the
velocity. We calculate the
KV shaking temperature $T^\nu_{sh}$\cite{koshvin} appropriate
to the drive and transverse directions, obtaining
it from 
\begin{equation}
T^\nu_{sh} =
\langle \sum_i^N (v_i^{\nu} - v^{\nu}_{cm})^2 \rangle /2N, ~~~~~~\nu = x,y.
\end{equation}
We measure the variation of the $T^\nu$'s as a function both
of force and $v_3$ at various points in our phase diagram.

The transition to the coexistence phase is marked by the growth of highly anisotropic, obliquely inclined
square nuclei, in a background of approximately triangularly coordinated
solid. To understand their role in the nucleation kinetics, we define and compute 
quantities which measure the anisotropy of such clusters. To determine
the source of  excess noise in the coexistence phase, we measure the probability distribution of the 
instantaneous velocity excess above the mean, as a function of the local coordination
as well as of the local density. As we show below, these measurements indicate that the substantial
portion of the noise originates from regions which are neither wholly 
square or triangular but to be found at the interface between such locally ordered
states.

\section{Phases and Phase Diagram of the Driven System}

A qualitative understanding of the basic structure of the phase diagram of 
Fig.~\ref{phasedia} can be
obtained from snapshots of instantaneous configurations in the steady state.
Such snapshots are shown in Fig.~\ref{Recryst}, obtained from simulations
at $T=0.1$ with $v_3 = 6.0$. 
The configurations labelled (a)-(f)
are   for driving forces $F_x  (= F) = 6, 10, 20, 22, 32$ and $40$ respectively. 
Particles are colored according to their 
coordination number $n$ = 4 (magenta), 5 (green), 6 (blue), 7 (orange)
and 8 (grey), obtained from Delaunay triangulations
and a Voronoi analysis\cite{preparata}. 

Disorder-induced inhomogeneities  spawn local defects and
dislocations ($5$ and $7$ coordinated particles) in the system
at low $F$. Boosting the external force  transforms the
disordered pinned phase, into a  plastically
flowing disordered fluid-like ``plastic-flow'' phase. This then turns into a  coherently 
moving triangular phase at still larger $F$.
\begin{figure*}
\begin{center}
\includegraphics[width=16.0cm,angle=0]{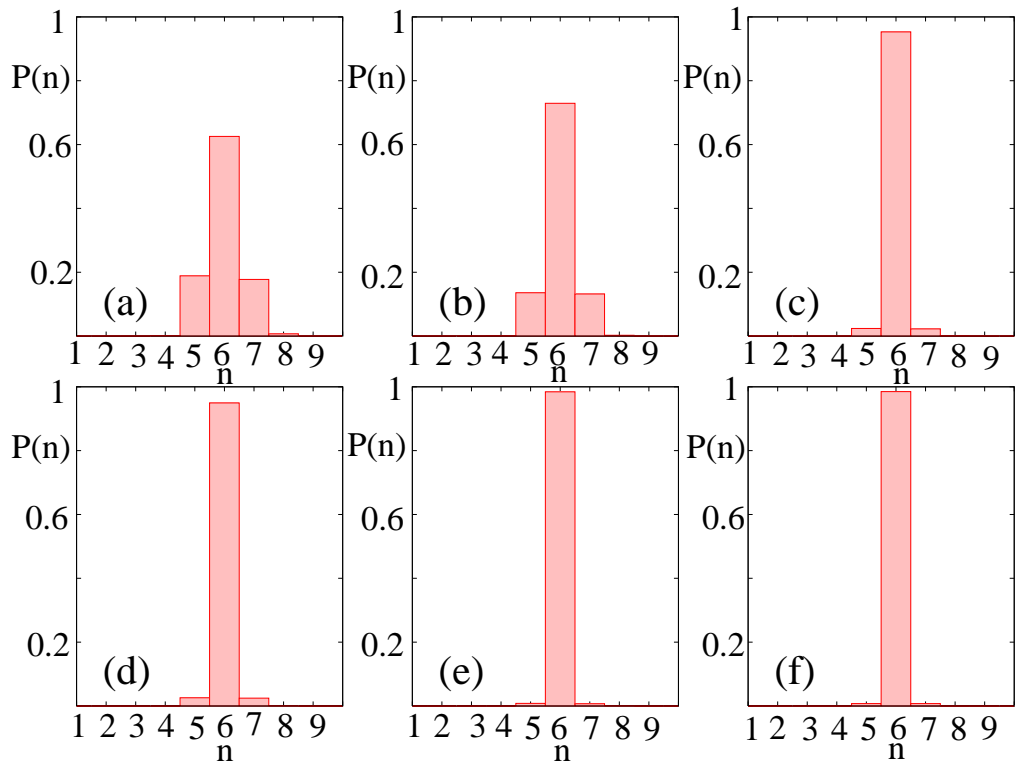}
\end{center}
\caption{
[Color Online]:The coordination number probability distribution at successive stages of 
freezing of the driven liquid, at driving forces $F_{x} = 6 (a), 10 (b), 
20 (c), 22 (d), 32 (e)$ and $40 (f)$ and with $v_3 = 6$. These forces correspond to
the forces given in the configuration snapshots of Fig.~\ref{Recryst}.
}
\label{CoordNo}
\end{figure*}
\begin{figure*}
\begin{center}
\includegraphics[width=12.0cm]{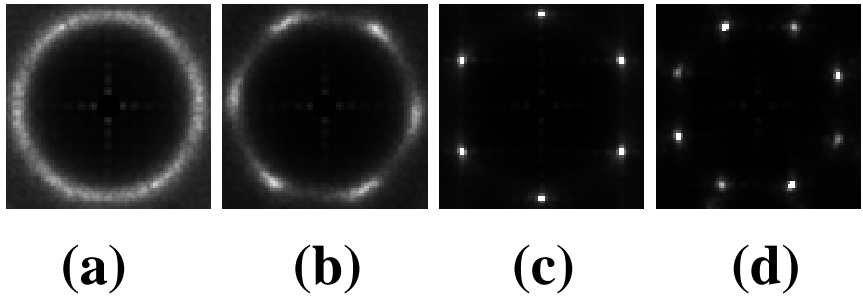}
\end{center}
\caption{
Structure factor 
$S({\bf q})$ for the plastic flow state
(a), anisotropic hexatic  phase(b), moving triangle (c) and 
moving square solid phases (d) at $v_3 = 6.0$ and $F_x = 10, 20,
60$ and $140$ respectively. To obtain $S({\bf q})$, $50$
independent configurations
were used. The structure in (d) reflects the presence of
two misoriented square crystallites.
}
\label{Sqfigs}
\end{figure*}

We show the evolution of
the coordination number histograms of the system for the corresponding forces
in Fig.~\ref{CoordNo}. In the disordered state, 5-fold and 7-fold coordinated
particles are roughly similar in number as expected. 
The number of particles with 5 and 7 fold coordination decrease with the drive
(Fig.~\ref{CoordNo} (a),(b)). There appears to be a small intermediate regime where
we see some clustering of these defects, as in (c) and (d). 
Finally, at much larger $F$,  the number of  non-hexagonal coordinated particles
declines abruptly (Fig.~\ref{CoordNo} (e),(f)) and the system freezes into a triangular latttice. 
Similar observations hold for the case of scans in $F$ at larger $v_3$, with the difference
that the ultimate large-force state is the flowing square lattice. Configurations in the ``coexistence
regime'' obtained at much larger forces ($F  \sim 100$ for $v_3 = 6$) are discussed separately in the sections 
which follow.

\subsection{Pinned Phase:}

For small $F$ the solid is pinned. In this regime, for
$F < F_c$, there is no centre of mass motion at the
longest times we simulate.  We  see
transient motion in the initial stages of the
application of the force, which then dies down once the
system optimizes its location within the background of
pinning sites.  
As $F$ is increased across the depinning
threshold, there are long transient time-scales for
motion to set in\cite{ankush1}. These time-scales appear to diverge as the system
approaches the transition, in agreement with expections
concerning continuous depinning transitions. However,
depinning appears to be hysteretic, on the length and time
scales we consider {\it i.e.} the reverse path, from depinned to
pinned states yields an abrupt transition between these 
states\cite{ankush, ankush1}. The depinned state just above the transition is inhomogeneous and undergoes
plastic flow\cite{jensen,jensen1,shiberlinsky} 
consistent with earlier numerical work.
For larger $F_x$ the velocity approaches the
asymptotic behaviour $v_{CM} = F_x$.

The pinned phase is a phase with short-range order in both translational and orientational
correlation functions. Correlations typically extend to about 4-8
interparticle spacings at the levels of disorder we consider.  At non-zero noise strength,
strictly speaking, we would expect an activated creep component
to the motion. This  presumably lies beyond the time scales of our 
simulation, given the relatively low temperatures ($T \ll T_m$) we work 
at.
\subsection{Plastic flow phase}

Upon increasing the force, we enter a regime of substantial
plastic flow\cite{jensen, jensen1,shiberlinsky}. This regime is one in which particle motion is extremely
inhomogeneous. Similar plastic flow is seen  in a large number of
simulations of particle motion in a random pinning background\cite{jensen,jensen1,shiberlinsky,koshvin,faleski,olsreinori,fangohr,chandran,kolton,otterlo}. What
is unusual here, however, is the strongly non-monotonic character of
the plastic flow boundary, as illustrated in the phase diagram of Fig.~\ref{phasedia}.
Note that the plastic flow phase boundary in the $F-v_3$ plane is largely independent of
$F$ for small $v_3$. However, once the critical value of $v_3$ for the square-triangular transition
is crossed, this boundary  becomes a  strong function of $F$, with the plastic flow region
expanding considerably before it collapses again. 

The structure factor $S(q)$ of the plastically
moving phase (C) obtained in a  narrow region just
above the depinning transition consists of
liquid-like isotropic rings, as shown in Fig.~\ref{Sqfigs}(a). 
For much larger values
of $v_3$, the disordered, pinned phase appears to depin directly
into the highly ordered moving square lattice phase, with no
trace of a plastic flow regime. We have searched for tetratic phases, with
algebraically decaying tetratic correlations
in the vicinity of this depinning transition. However, no such phase is  apparent
in our numerics.

In the plastic flow regime, transport properties are noisy, reflecting the 
underlying highly disordered nature of the phase, in agreement with previous 
work on such plastic flow states in systems without competing phases\cite{jensen,jensen1,shiberlinsky,koshvin,faleski,olsreinori,fangohr,chandran,kolton,otterlo}.
We
see no evidence for a two-peak structure in the velocity distribution function,
unlike some previous work, provided we average enough. Thus no particle is
stationary over the full time-scales of our simulations. 

\subsection{Anisotropic Hexatic Phase}

As $F$ is increased further, we encounter a narrow regime in which translational correlations
are short-ranged while angle-averaged orientational correlations appear to
decay as power laws. Within this phase the circular 
ring in $S(q)$ concentrates into
six smeared peaks, as shown in Fig.~\ref{Sqfigs}(b). The presence of 6-fold order
in the absence of the sharp Bragg-peaks associated with 
crystalline ordering suggests that this phase may have
hexatic orientational order.  We thus tentatively 
identify this phase as a driven hexatic, as shown in  (D)\cite{dpt-hexatic,otterlo};
the terminology ``anisotropic'' is justified in what follows. 
However, to strengthen this assignment, other possible 
assignments, such as to a coexistence regime known to
plague similar analyses of hexatics in two dimensional systems,
must be ruled out\cite{jaster}.

\begin{figure}
\begin{center}
\includegraphics[width=8.0cm]{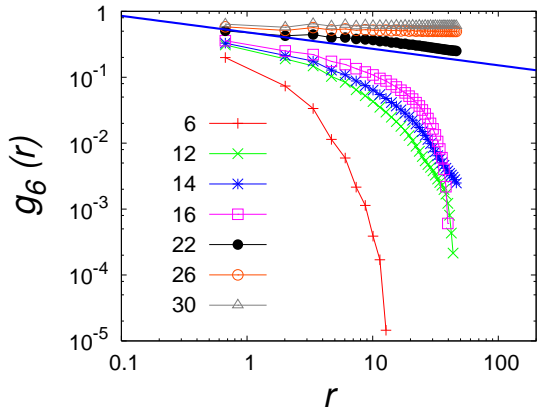}
\end{center}
\caption{[color Online]:
The hexatic correlation function $g_6(r)$ for $F = 6,12,14,16,22,26$
and $30$, each averaged over 50 independent configurations of a $N=10000 $
particle system at   $v_3=6$. The solid line indicates the universal behavior 
$g_6(r)\sim r^{-1/4}$ at the liquid to hexatic transition. 
}
\label{HexCorr}
\end{figure}

We compute the correlations of the hexatic order parameter
for a very large system at varying values of $F$, as shown in 
Fig.~\ref{HexCorr}, where $N=10000$. This figure displays
the evolution of  $g_6(r)$, as $F$ is
varied across the phases $(C) \to (D) 
\to (E)$ at fixed $v_3 = 6$. We observe a sharp exponential
decay of hexatic correlations in (C) and find that at the liquid to hexatic
transition  for $F \simeq 22$, the decay fits
the universal behavior 
\begin{equation}
g_6(r)\sim \frac{1}{r^{1/4}}
\end{equation}
expected at the fluid hexatic
transition in two-dimensional non-disordered fluids. This exponent is obtained
in a relatively narrow range of forces. Our system sizes  are comparable to
typical sizes employed
to observe a metastable hexatic phase in two dimensional melting of pure
solids.
\begin{figure*}
\begin{center}
\includegraphics[width=12.0cm]{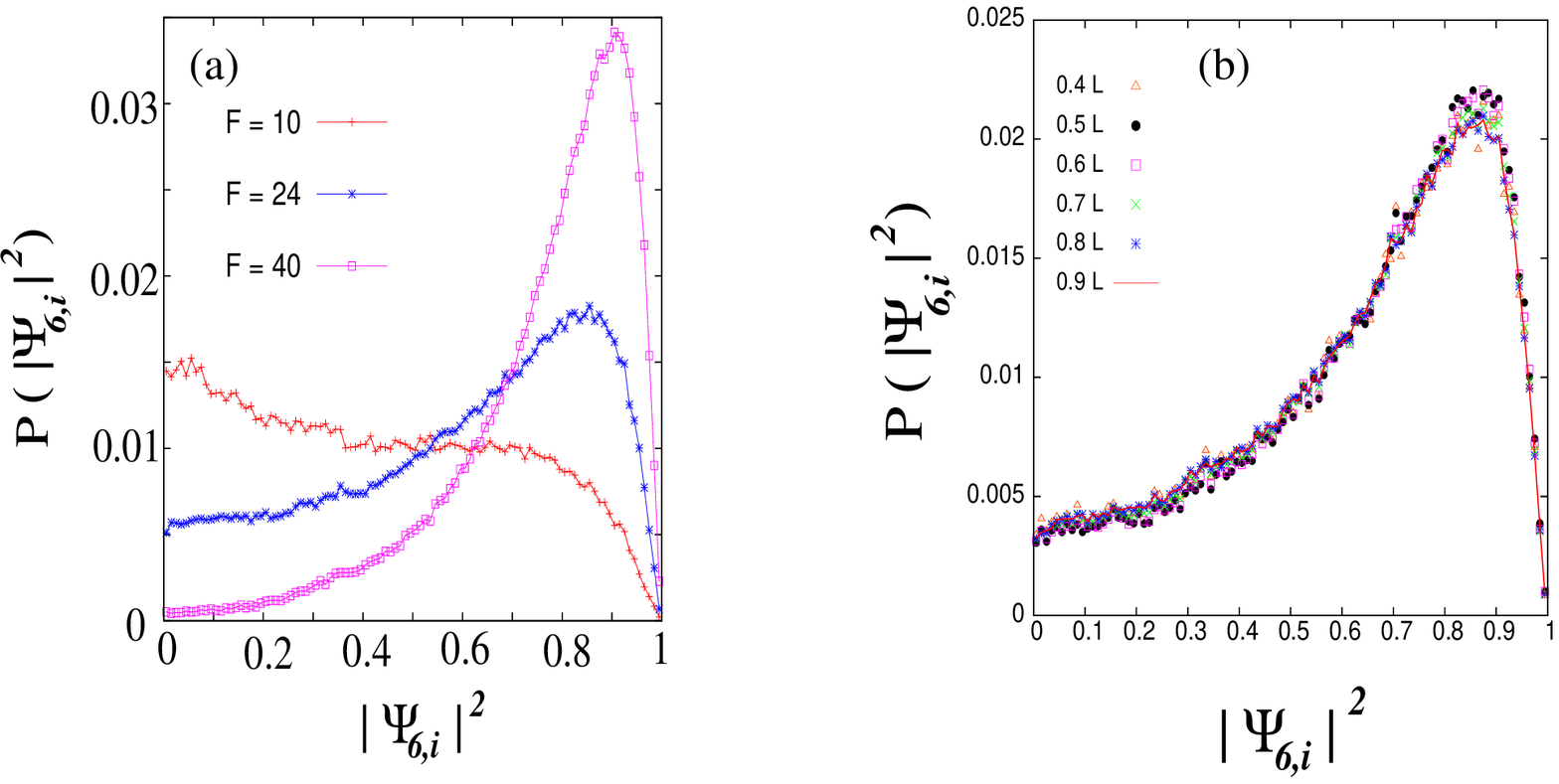}
\end{center}
\caption{[Color Online:]
(a)The distribution of the amplitude of the local hexatic order parameter $|\psi_{6,i}|^2$ 
over the system for $N = 10000$ in the plastic ($F_x = 10$), hexatic-glass ($F_x = 24$) and
moving triangular ($F_x = 40$) regimes, with $v_3 = 6.0$
(b)  Scaling of the distribution of $|\psi_{6,i}|^2$
deep in the hexatic regime for $F_x = 26$, computed over fractions $0.4, 0.5,
0.6, 0.7, 0.8$ and $0.9$ of the simulation box length $L$, with $v_3 = 6.0$.
}
\label{combinedscaling}
\end{figure*}

In Fig.~\ref{combinedscaling}(a), we compare the distribution of $|\psi_{6,i}|^2$
over the full system, for the plastically flowing, moving triangular
and the intermediate hexatic-glass regimes. The figure illustrates that in
the liquid/plastic regime ($F=10$),  the distribution is primarily
governed by non six-fold coordinated particles. In the high
driving force regime ($F=40)$, six-fold coordinated particles 
contribute in main. This raises the question of whether the intervening `hexatic'
regime we observe in terms of $S({\bf q})$ and the correlation of the
hexatic order parameter $g_6(r)$ at $F=24$, is truly hexatic or defined by a
coexistence of the `solid' and `disordered' phases. If configurations resemble
those at  solid-fluid coexistence, the distribution would be expected to be 
a sum of the disordered, relatively ordered  and interface distributions (weighted 
with their relative areas).

To settle this issue we investigate the dependence of the distribution
of $|\psi_{6,i}|^2$ on the size of the system. The distribution is obtained
by dividing the system into a number of blocks and computing the
distribution of  $|\psi_{6,i}|^2$ within each block for every configuration.
In this way, information over many length scales can be obtained from the
same set of configurations. This provides us with the distribution of $|\psi_{6,i}|^2$
for various fractions of the total system.

We find,  as shown in Fig.~\ref{combinedscaling}(b), that apart from
finite size effects, there
is not much difference between the distributions. Also, the distributions for
the two largest systems, {\it viz.} $0.8 L$ and $0.9 L$ ($L$ being the size of
the simulation box), coincide to within statistical errors. In the case of 
two-phase coexistence one would expect a strong size
dependence when the size of the blocks is comparable to the size of
the phase separating clusters. These observations
rule out  two-phase coexistence, at least of the conventional kind,
as an explanation for the long orientational
correlations obtained on circular
averaging.

As we show below, a  more detailed characterization of this phase
 yields the following results: Long-range order in orientation is always present in the
direction of drive, while remaining short-range in the perpendicular 
direction. When averaged over all orientations, the resulting function
appears power-law-like over 2-3 decades {\it i.e.} over length scales addressed in
virtually all simulations so far. Thus, while we 
call the intervening phase as the 
``anisotropic hexatic'', we diverge sharply from previous work in
our claim  that the true phase is never a true hexatic in the real sense, since it
always has long-range order in the drive direction. We conclude 
that the apparent hexatic correlation arises, in fact, as an artefact of circularly averaging
a very anisotropic correlation function, with qualitatively different decays in the 
longitudinal and transverse directions.

\subsubsection{Analogy to the XY model}
The power-law decay of orientational correlations shown in 
Fig.~\ref{HexCorr}, taken together with the scaling of $|\psi_{6,i}|^2$
suggests that the driven 
system in the vicinity of $F= 22$ might best be described as a hexatic,
with power-law correlations in the local orientation but short-range 
translational order. Such a state is 
analogous to the low-temperature phase of the 2-dimensional XY
model, where the Mermin-Wagner theorem rules out long-range order
at any finite temperature but vortex-like excitations responsible for the
transition to the disordered phase exist chiefly as bound vortex-anti-vortex 
pairs.  However,  the presence of  the drive direction introduces an anisotropy into the system. The
consequences of this anisotropy have not, to the best of our
knowledge, been explored in previous studies of putative hexatic
phases in driven disordered solids\cite{dpt-hexatic}. 

In our studies, we have been motivated by an analogy to the physics of the 
XY model in an applied field,  specifically by the intriguing possibility that the drive  
might play a role equivalent to that 
of the magnetic field in the XY model. Renormalization group studies~\cite{fertig} of the 2-dimensional 
XY model in an external magnetic field with Hamiltonian, 
\begin{equation}
\mathcal{H}_{XY} = \sum_{i,j} J_{ij} s_i s_j \cos(\theta_i - \theta_j)
- h\sum_i s_i \cos(\theta_i)
\end{equation}
where the external field $\bf h$ points along the positive $x$-axis,
indicate three distinct phases.
These are a linearly confined phase, a logarithmically confined phase and a
free vortex phase obtained as the temperature is gradually increased. 
Vortex-antivortex pairs at low temperatures are linearly confined
by a string of overturned spins. With increasing temperature these
strings participate in a proliferation transition, but  vortices
still remain confined due to a residual logarithmic attraction. As
the temperature is further increased, the vortices overcome this
attraction and  are finally deconfined.  

Since the external field breaks rotational symmetry, the  magnetization
is  non-zero in all the three phases.
Further, the free energy is  argued to be smooth; thus
the distinction between phases can only be found in the structure and
distribution of topological defects.
The correlation function $\langle \cos(\theta_i)\cos(\theta_j) \rangle$
is then predicted to saturate to a constant value ($b$, say) for
$|{\bf r}_i - {\bf r}_j| \to \infty$.
This constant $b \propto h^2$ since the order parameter ($\cos(\theta_i)$)
couples linearly  to the field $h$.
On the other hand, the correlation function
$\langle \sin(\theta_i)\sin(\theta_j) \rangle$ which corresponds to
spin fluctuations in the direction transverse to the applied field
(see Eq. (6.6)), is predicted to show an exponential decay in all the three phases.
\begin{figure*}
\begin{center}
\includegraphics[width=10.0cm]{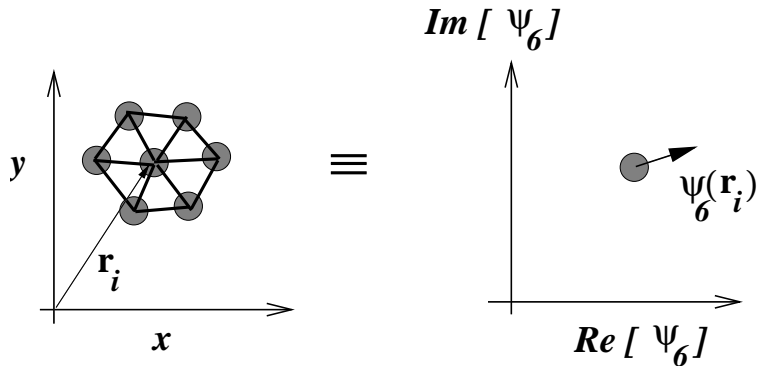}
\end{center}
\caption{
Mapping to the XY model : local hexagonal configuration around the
particle $i$ (diagram on left) may be analyzed in terms of the phasor
$\psi_{6}({\bf r}_i)$ represented as a spin situated at ${\bf r}_i$
(diagram on right).
}
\label{MapSpin}
\end{figure*}
\begin{figure*}
\begin{center}
\includegraphics[width=18.5cm]{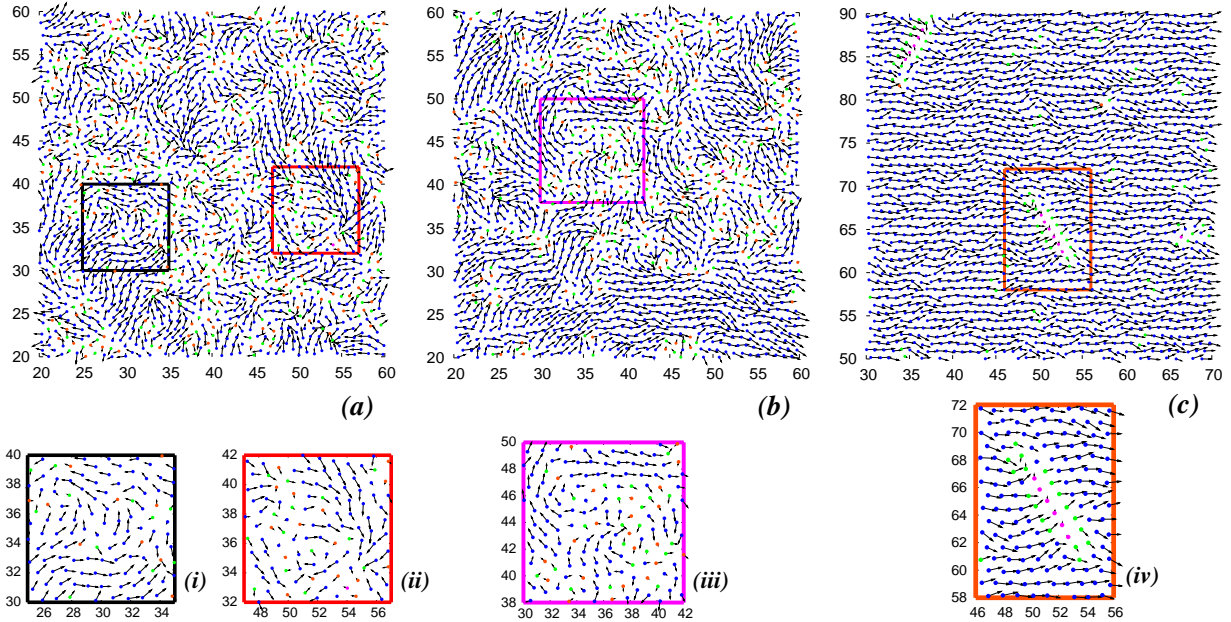}
\end{center}
\caption{[Color Online]:
Typical orientation of the `hexatic-spins' (see text) associated with
particles in the liquid (a), anisotropic-hexatic (b) and triangular (c)
phases, at the driving forces $F_{x} = 10, 24$ and $40$ respectively.
The boxes in (a), (b) and
(c) are expanded in (i) and (ii) (corresponding to the configurations boxed in
(a)), in (iii) (corresponding to the configuration boxed in (b)) and
(iv), corresponding to the boxed configuration in (c), to enable an easier visualization of spin configurations.
The colors correspond to local coordination, $n = 6$ (blue), $5$ (green),
$4$ (magenta) and $7$ (orange). Note the presence of  extended defects
in the ordering shown in (c),  associated with ``strings'' with local square symmetry. In all the three 
pictures only a part of the simulation box is shown for the sake of clarity 
and the configurations are obtained for $v_{3} = 6$. 
}
\label{HexSpin}
\end{figure*}

\subsubsection{Quantifying Local Orientational Order}
To test this possibility, we must  map local orientational order to an 
appropriate XY-like two-dimensional vector from which defect structures can be
extracted.  This  is  done through the 
local hexatic order parameter, $\psi_{6,i} \equiv \psi_{6}({\bf r}_i)$,
defined earlier. It is  convenient to identify this local quantity (a phasor) with a
`hexatic-spin' with components $Re[\psi_{6,i}]$ and $Im[\psi_{6,i}]$, thus mapping 
local geometrical order to a soft-spin XY degree of freedom. 

We illustrate this mapping from the local orientations of particles in
real  space to XY spins in Fig.~\ref{MapSpin}.
Unlike in  the XY model, such hexatic spins
are not attached to a fixed lattice but  associated with the 
moving particles.  They thus encode important information 
concerning  the equal-time orientational correlations
in the driven system.  Fig.~\ref{HexSpin} shows 
real space configurations of the particle system with associated
hexatic spins at various values of $F$, illustrating the variety
of associated spin configurations present in this mapping. 

These configuration maps enable the identification of topological defects
in the ordering in the associated XY model.  Note the presence of locally aligned regions as
well as vortex-like excitations of strength 1 and 1/2.   Qualitatively,  we find that in both the disordered  
liquid phase and the anisotropic hexatic phases, vortex configurations in 
such spin configurations appear to have little correlation with each other, suggesting that they may be either
unbound or relatively weakly bound at best (see Fig.~\ref{HexSpin} (a) and (b)); the boxed regions of
these figures as indicated are expanded in Fig.~\ref{HexSpin}(i) and (ii) for (a) and in (iii) for (b). 
However, we have not  been able to establish a quantitative distinction between the disordered liquid phase 
and the anisotropic hexatic phase using our simulation data. 

\begin{figure}
\begin{center}
\includegraphics[width=8.0cm]{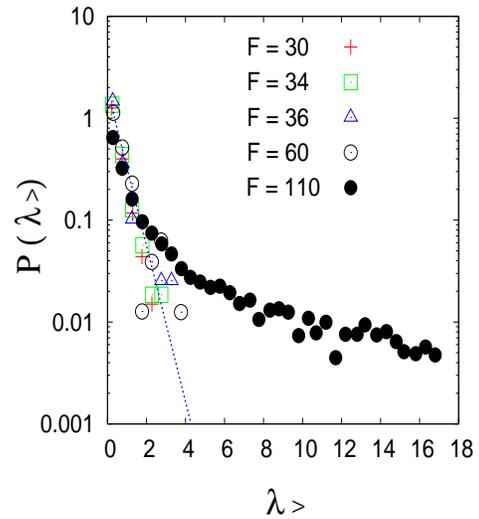}
\end{center}
\caption{[Color Online]:Probability distribution of the largest eigenvalue $\lambda_>$ in a semi-logarithmic scale for external forces $F = 30, 34, 36, 60$ and $110$. It is clear that for all but the largest force, the probability distribution is exponential showing that the energy of string-like excitations within the triangular phase is linear with size. For the largest force which is inside the coexistence region (see text), the probability distribution decays slower than an exponential.}
\label{string}
\end{figure}

Some quantification is, however, possible for larger forces, in the flowing  triangular phase,
(Fig.~\ref{HexSpin} (c)), where defects in the ordering appear to be associated with strips, or strings, of defect. In this
case, the defect is locally a solid with square symmetry. (An expanded plot of the boxed region in Fig.~\ref{HexSpin}(c) is
shown in (iv), illustrating the one-dimensional character of the defect). We expect that such defects should be 
{\it linearly} bound, with an energy proportional to the length of the
strip. The binding energy in the latter case should be proportional to a non-equilibrium analog of a 
surface tension between the square and the triangular crystal. 
This argument is supported by calculations of the moment of inertia tensor of the set of particles  which belong to such a defect,
averaged over configurations which contain such defects. The largest eigenvalue ($\lambda_>$) of this tensor measures the length of these 
extended string-like defects. 
In Fig.~\ref{string} we show the probability distribution $P( \lambda_> )$ at a few different values of the external drive. It is clear that within the intermediate
force regime, the 
distribution is exponential,  implying that the energy for these excitations scale linearly with their size. These string like excitations align
preferentially along the crystallographic axes of the surrounding triangular lattice. As the force is increased, these defects offer nucleation sites for 
square crystals which are less anisotropic. The probability distribution then ceases to be linear in $ \lambda_>$, leading to the long tail in the data shown in
Fig.~\ref{string} at $F=110$ . 

\begin{figure*}
\begin{center}
\includegraphics[width=18.0cm]{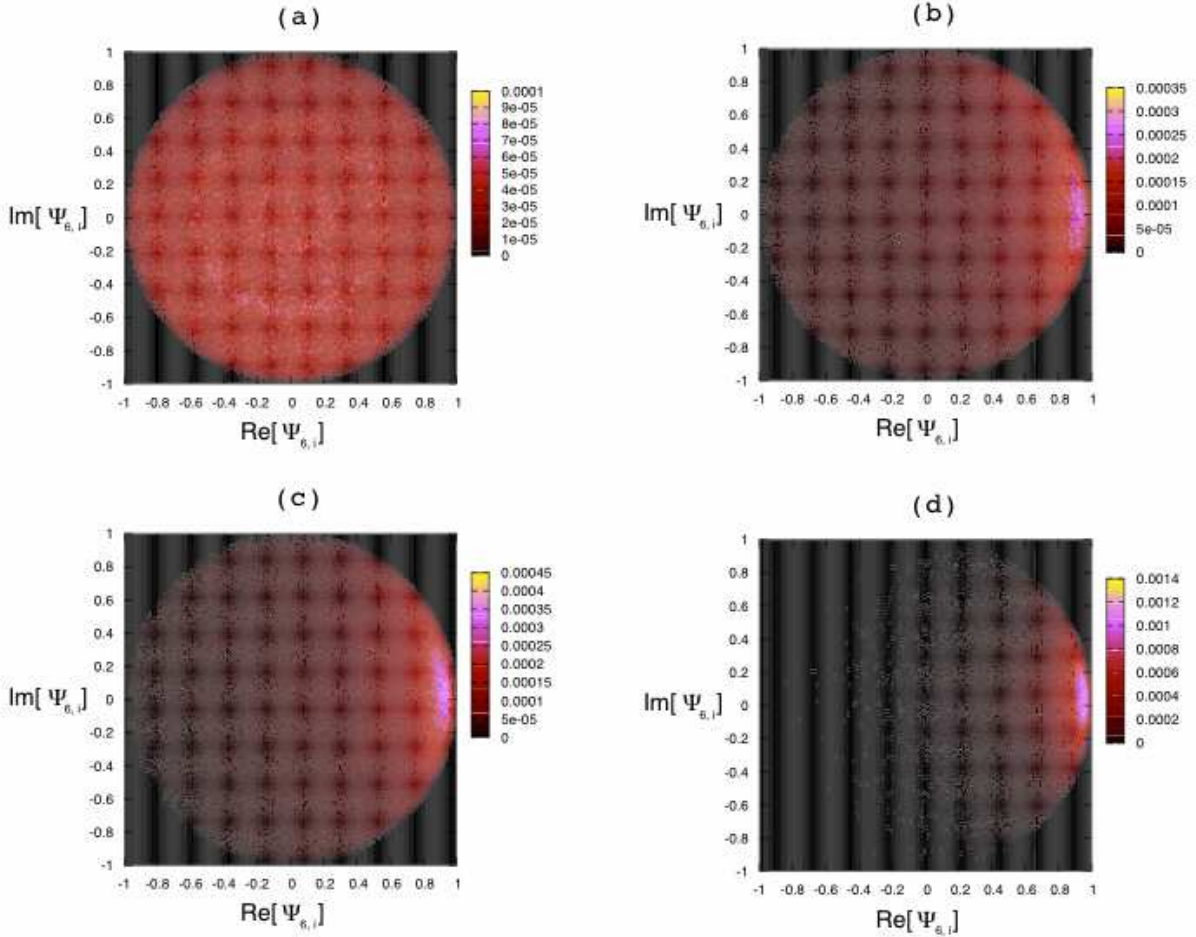}
\end{center}
\caption{[Color Online]:
Probability distribution of the `hexatic-spin' phasors for forces
$F_x = 8$ (a), $22$ (b), $24$ (c)
and $40$ (d). The plane of the Argand Diagram within the
area enclosed by the points $(-1,-1)$, $(1,-1)$, $(1,1)$ and $(-1,1)$
is subdivided into $200 \times 200$ square boxes (of width $0.01$).
The sum of the probability distribution over these boxes is normalized
to unity.
}
\label{Argands}
\end{figure*}

Fig.~\ref{Argands} shows the probability distribution of the `hexatic-spin' phasors for forces
$F_x = 8$ (a), $22$ (b), $24$ (c)
and $40$ (d). While Fig.~\ref{Argands}(a), obtained within the plastic flow phase,
appears to have a uniform distribution of hexatic spins with angle, Figs.~\ref{Argands}(b), (c) and
(d) display substantial non-uniformity in this distribution.
Note the following feature of Fig.~\ref{Argands}(c): The 
mapped spins tend to overwhelmingly point along the drive direction
in the flowing triangular state. This should be 
contrasted with the fact that  the corresponding distribution function for the X-Y model in 
zero external field is isotropic across the disordered to 
quasi-long-range-ordered  transition, peaking below it at
$\vert \Psi_6 \vert \neq 0$, a value  independent  of the phase angle, in a finite
system.  This indicates  that orientational order in our problem is strongly biased
by the drive $F$, if $F$ is sufficiently large.  
\begin{figure*}
\begin{center}
\includegraphics[width=14.0cm]{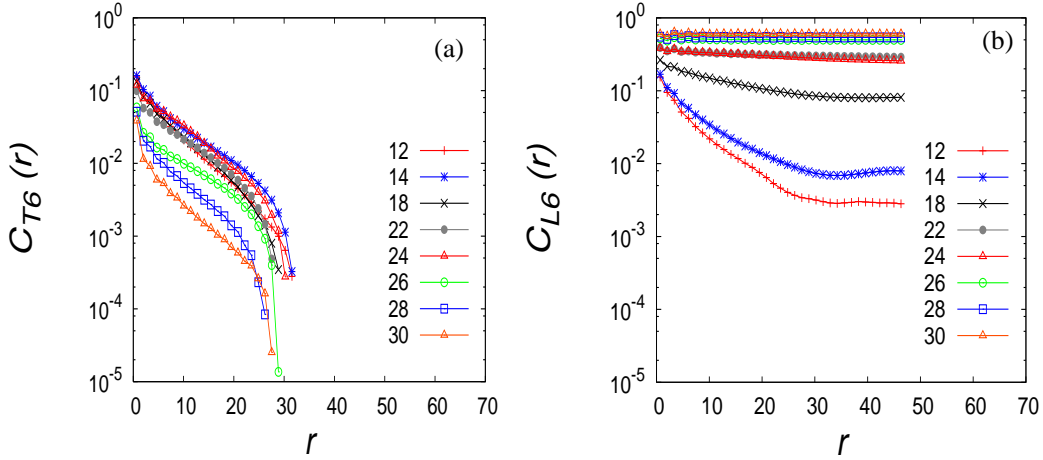}
\end{center}
\caption{[Color Online]:
(a) The correlation functions $C_{T6}(r)$ plotted 
for a range of force values $F_{x} = 12,14,18,20,22,24,26,28$ and $30$ and for
$v_3 = 6.0$. Note the
generic exponential decay obtained. (b) The correlation function $C_{L6}(r)$ (see text) plotted 
for a range of force values $F_{x} = 12,14,18,20,22,24,26,28$ and $30$.
in Ref.\cite{fertig}.}
\label{combined}
\end{figure*}
The probability distribution of the hexatic-spins in the complex Argand
plane formed by the Re$[\psi_{6,i}]$ and Im$[\psi_{6,i}]$ axes 
clearly describes how the external symmetry-breaking
field ($F_x$) builds up anisotropy in the our system, thus ordering the
spin orientations. At low drive,  the anisotropy is masked by
disorder-induced fluctuations at the scale of our simulation box. As such fluctuations are increasingly
suppressed at higher
drive values, the system appears to organize into a coherently moving lattice structure whose principal
axes are biased by the  force. 
\begin{figure}
\begin{center}
\includegraphics[width=8.0cm]{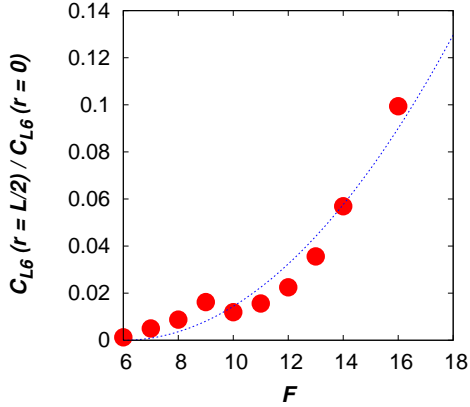}
\end{center}
\caption{[Color Online:]
The constant part in the angular correlation function along the drive
direction goes quadratically with the driving field strength above $F_c$
as predicted by Fertig {\it et al.}\cite{fertig}.
The data points (circles) are for
$v_3 = 6.0$ and the dashed line is a guide to the eye.
}
\label{SpecOP}
\end{figure}
The spin-spin correlation functions 
in our model, Fig.~\ref{combined}(a) and Fig.~\ref{combined}(b), obtained via
the correlation functions $C_{L6}(r)$ (longitudinal) and 
$C_{T6}(r)$ (transverse), correspond to the local quantities 
Re$[\psi_{6,i}]$ and Im$[\psi_{6,i}]$. They are
\begin{equation}
C_{L6}(r) = \langle  Re[\psi_{6}(0)] Re[\psi_{6}(r)]   \rangle
\end{equation}
\begin{equation}
C_{T6}(r) = \langle  Im[\psi_{6}(0)] Im[\psi_{6}(r)]   \rangle
\end{equation}
In all the phases (C), (D) and (E) of Fig.~\ref{phasedia}, the 
orientational correlation function $C_{L6}(r)$  saturates asymptotically,
as expected, to a constant. In phase (D), such
saturation is obtained only in the drive direction. The transition
from phase (D) to phase (E), when triangular {\it translational} order 
increases continuously with $F_x$, appears to be smooth. The correlation function
$C_{T6}(r)$ decays exponentially  in all the three phases.

Note that our mapping for orientational order and its correlations in the
moving state  is
free from any underlying lattice effects, in contrast to earlier studies of the XY model
in a field. Therefore, any anisotropy
reflected in the spin distribution (see Fig.~\ref{Argands}) or their
correlation functions reflects an intrinsic property of the
particular driven phase. The quadratic dependence
of the asymptotic saturation value of $C_{L6}(r)$ on the applied
field $F_x$ within the fluid (plastic flow) regime, as predicted by this intuitive mapping, 
is shown in Fig.~\ref{SpecOP}.

The results presented here address one fundamental issue in the 
literature on orientational order in driven disordered states, in
particular the question of whether a quasi-long-range ordered
phase, the ``hexatic'' can exist. We show here conclusively that
it cannot. What exists is an unusual intermediate
state which possesses long-range orientational order in the direction singled
out by the drive whereas orientational order decays exponentially
in the transverse direction.

\subsection{Phases at Large Force: Square, Triangular and  Coexistence}

What happens at still larger force values depends on the value of
$v_3$. For small $v_3$, the system transits directly from anisotopic
hexatic glass to triangular moving crystal. The Bragg peaks sharpen into
sharp Bragg spots with six-fold symmetry. At intermediate values of $v_3$, 
the system appears to undergo
an unusual transition into what we  term a ``coexistence
phase'', discussed in more detail below. In this phase, the system has both triangular and 
square domains and interconverts between them over a broad
distribution of time-scales.
Inspection of configurations suggests an analogy to equilibrium
phase coexistence with a large, heterogeneous distribution of domain
sizes of triangular and square regions, although this is an explicitly non-equilibrium system.

At larger values of $v_3$, the coexistence regime appears bounded. However,
the disordered, plastic flow regime is observed to expand. At these
values of $v_3$ the system undergoes a direct transition into the
square phase. For much larger $v_3$, the phase boundary between
plastic and square phases collapses again with increasing $v_3$, reducing the extent
of the plastic regime. In this large $v_3$ regime, the system depins discontinuously and 
elastically from a 
pinned to moving square crystal with no intervening plastic flow phase that our
numerics can resolve.  The plastic flow regime 
(C), as well as that of the hexatic glass (D) expands at larger
$v_3$ due to the frustration of local triangular 
translational order by three-body interactions.  On further increasing $F_x$, the
structure obtained depends on the value of $v_3$: for low $v_3$
the final crystal is triangular (E) whereas for
large $v_3$ it is square (F). 
We assign these states through a study of the structure
factor $S(q)$, as well as the coordination number probability distributions shown in 
Fig.~\ref{CoordNo}, observing 
that the slighly smeared six-fold
coordination of the hexatic glass consolidates into sharp
Bragg-peaks across the transition into the ordered states, as shown in Fig.~\ref{Sqfigs}.

\subsection{The Coexistence phase}
For intermediate $v_3$ and $F$, the system exhibits a
remarkable ``coexistence'' regime (G)
best described as a mosaic of dynamically fluctuating
square and triangular regions. From a direct calculation  of the
structure factor we see the simultaneous appearance of
peaks corresponding to hexagonal and square order\cite{ankush1}.
The intensity of the peaks from the hexagonal and square phases are
comparable. 

\begin{figure*}
\begin{center}
\includegraphics[width=18.5cm]{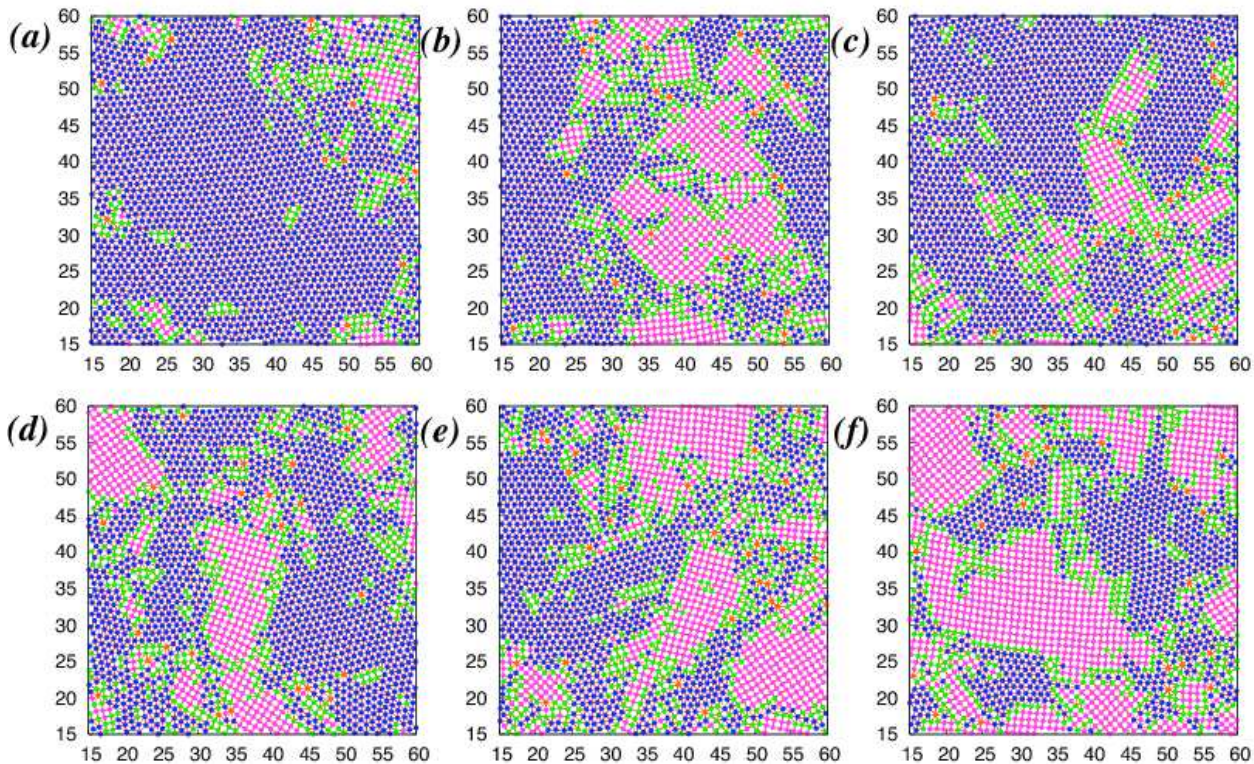}
\end{center}
\caption{[Color Online:]
Evolution of configurations
showing square-triangle coexistence with increasing force
$(a) F=90,(b) 95,(c) 100,(d) 105,(e) 110,$ and $(f)115$,
together with the computed 
Delaunay mesh. The particles are 
colored according to the number of neighbors n = 4 (magenta), 5 (green),
6 (blue) and 7 (orange). Particles with coordination 5 
are present mainly in the interfacial region while those with 7 are
associated with isolated dislocations.
}
\label{Coex}
\end{figure*}

As $F$ is increased, clusters of  $4$-coordinated particles
grow in size.  The evolution of the configuration with increasing force
is shown in Fig.~\ref{Coex}, which illustrates the square-triangle domain
mosaic characteristic of the coexistence regime. The $6$-coordinated regions decrease in size
and the dynamics of the interfacial region - with predominantly $5$-coordinated
particles and with a few isolated $7$-coordinated particles -  shows enhanced and 
co-ordinated fluctuations. Real space configurations (Fig.~\ref{Coex}) exhibit islands
of square and triangular coordination connected by interfacial regions with predominately 5
coordinated particles. The
configuration, as viewed in the co-moving frame, is extremely
dynamic, with  islands rapidly interconverting
between square and triangle. This interconversion
has complex temporal attributes.
\begin{figure}
\begin{center}
\includegraphics[width=8.5cm]{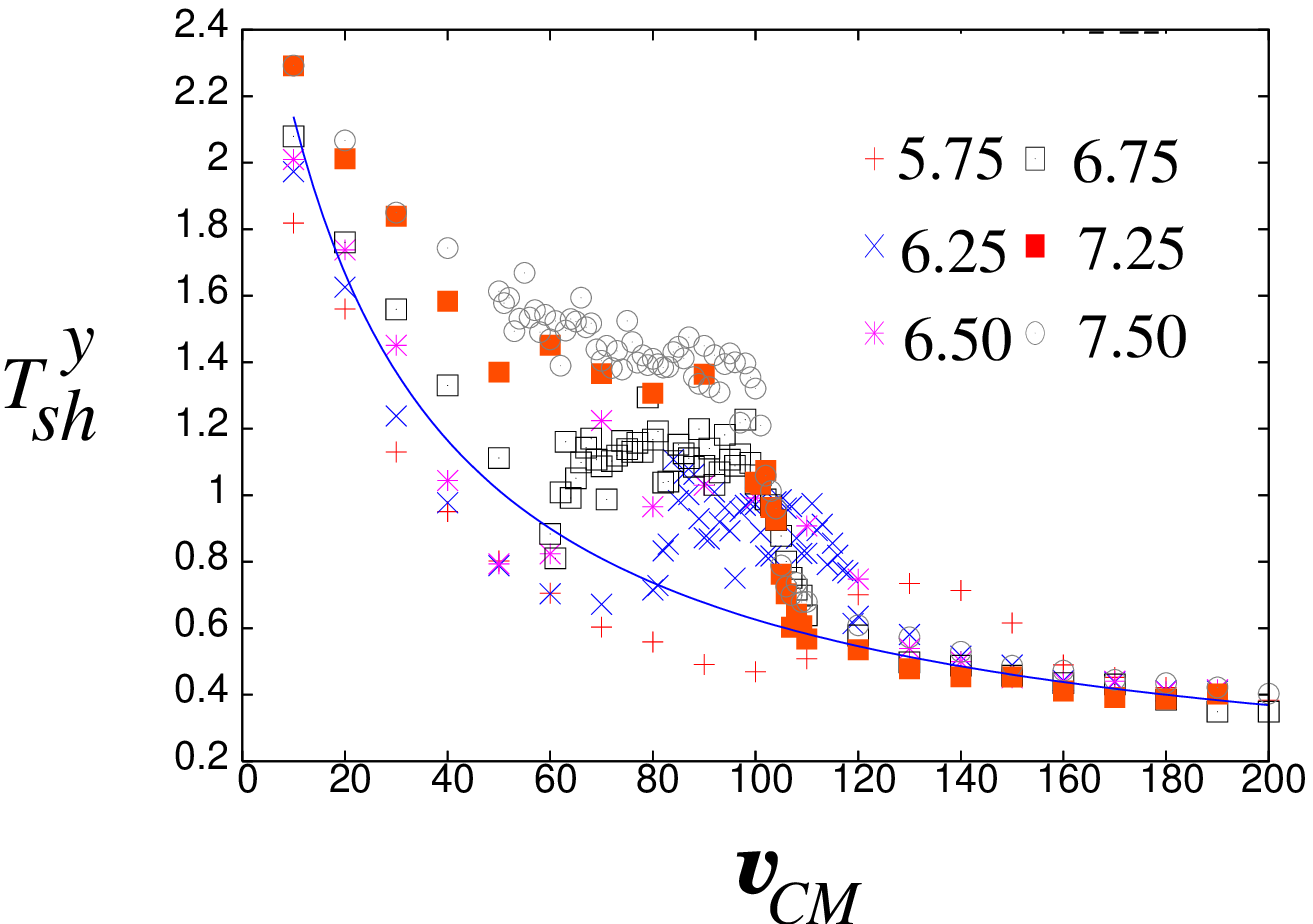}
\includegraphics[width=8.5cm]{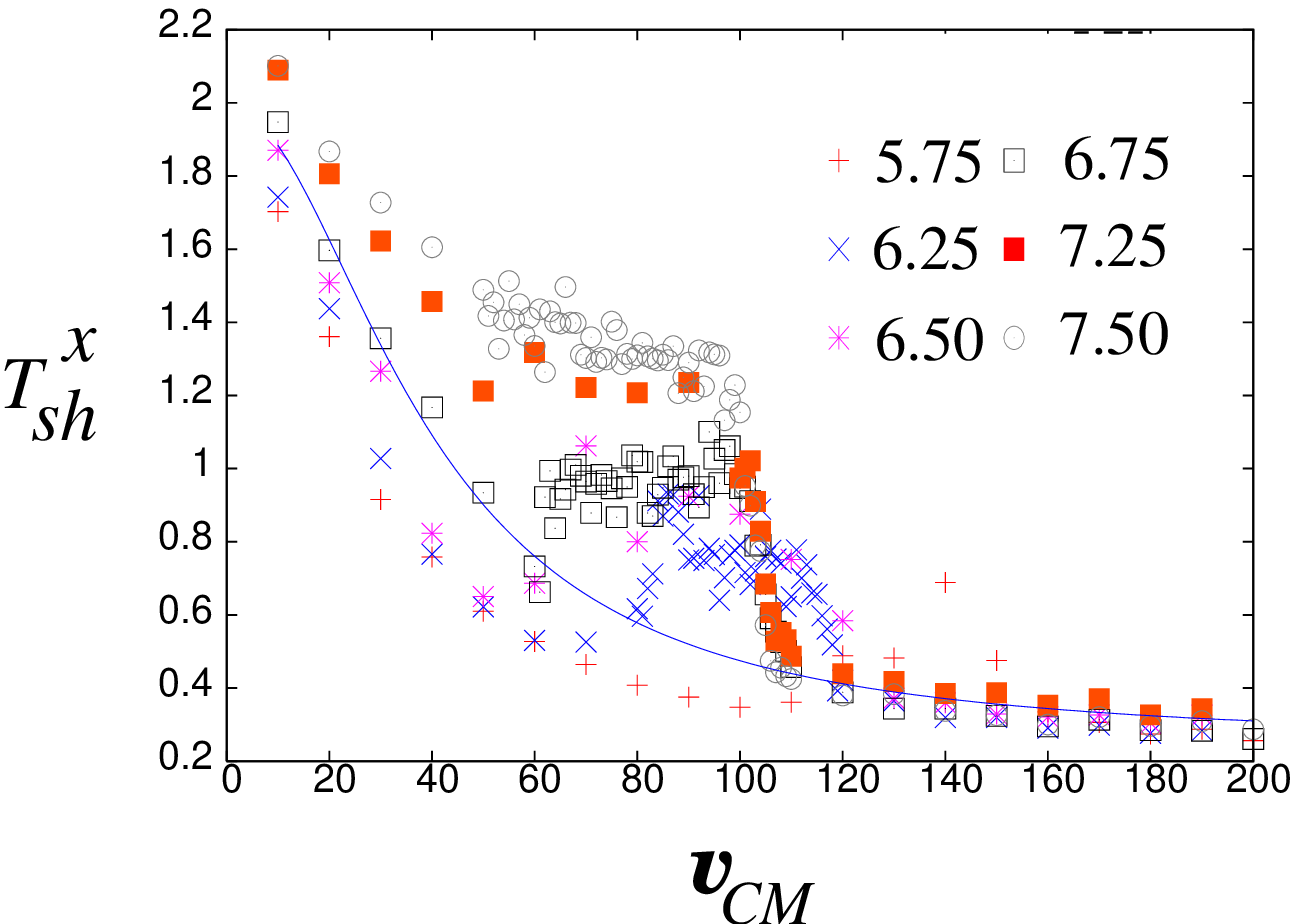}
\end{center}
\caption{[Color Online]:
The shaking temperatures in the transverse (top) and drive
(bottom) directions ($T_{sh}^{y}$ and $T_{sh}^{x}$) for different
$v_3$ as a function of the centre of mass velocity ($v_{CM}$), showing
the quantity is almost independent
of $v_3$ except in and around the coexistence regime, due to effects of
fluctuation.
}
\label{TempsXY}
\end{figure}

As  substrate randomness is averaged out
due to the motion of the particles,
the shaking temperature of the system can reasonably be expected to decrease. This is reflected in the
decrease of the width in velocity component distributions.
The ``shaking temperature'' predictions of Koshelev and Vinokur,
would indicate a $\sim 1/v$ and $\sim 1/v^2$ nature of the fall
in the transverse and drive directions respectively. Our results for $T^\nu_{sh}$
are in agreement with this prediction {\em outside} the coexistence regime.

We find, as shown in Fig.~\ref{TempsXY},
that $T^\nu_{sh}$ is nearly independent of $v_3$. However,
within the  coexistence regime,
$T^\nu_{sh}$ behaves non-monotonically. 
Typically, for a particular disorder configuration
and for $5.5 < v_3 < 8.5$, $T^\nu_{sh}$ appears to increase
sharply at a well defined $F_x$, signifying the start of
coexistence. Within G, $T^\nu_{sh}$ remains high but drops
sharply at the upper limit of G, to {\it continue to
follow the interrupted KV behavior}. This anomalous enhancement
of fluctuation magnitudes provides strong evidence
for a genuine coexistence phase, since increasing
the driving force would be expected to {\em reduce}
current noise monotonically once the system depins,
as observed in all previous simulation work on related
models \cite{faleski,fangohr}. The limits of
the coexistence region, though sharp for any typical
disorder realization, vary considerably {\it between}
realizations.

\begin{figure*}
\begin{center}
\includegraphics[width=18.5cm]{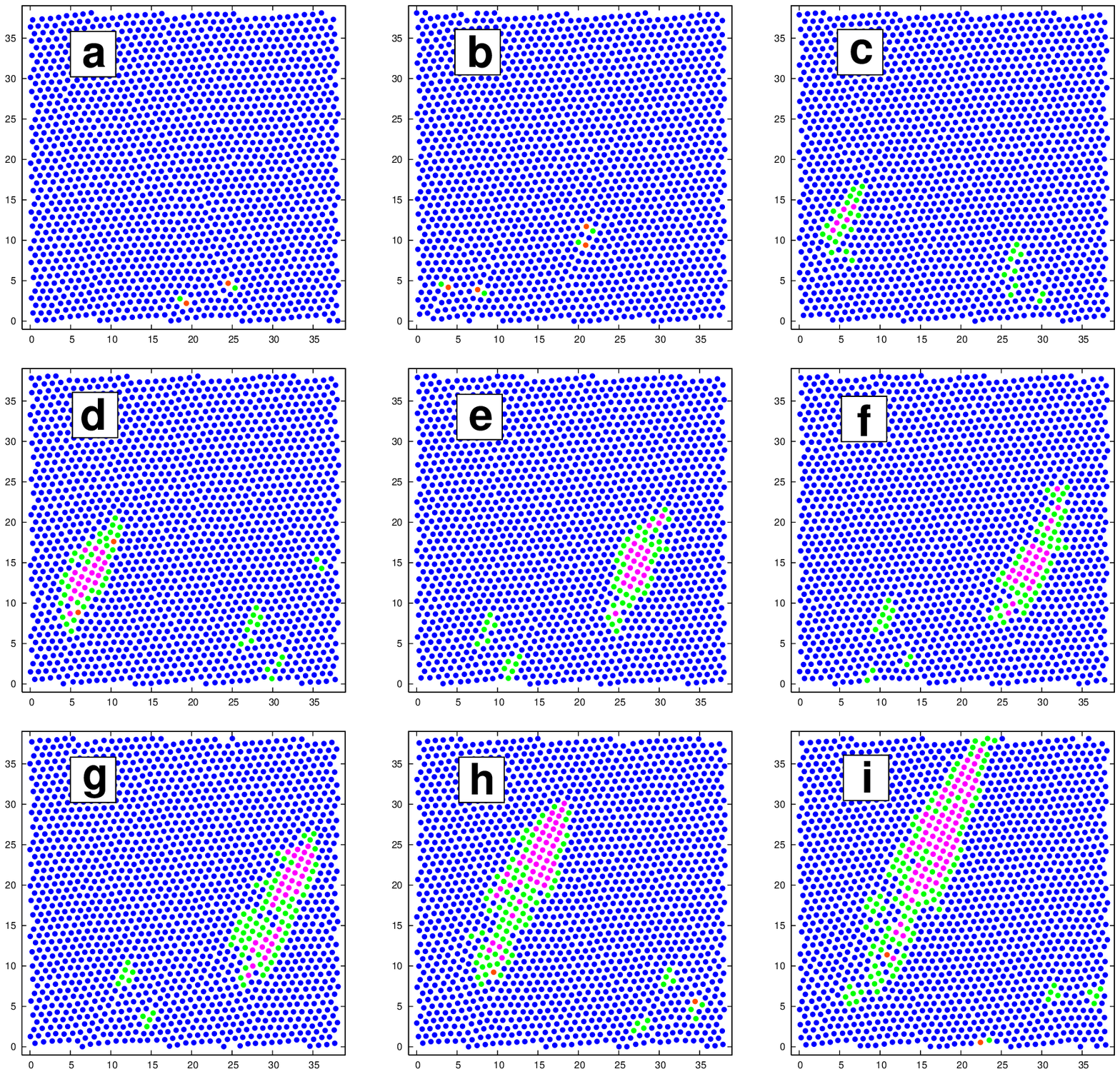}
\end{center}
\caption{[Color Online:] Dynamical nucleation of a square defect against a triangular background,
shown for $F= 98$, $v_3 = 6.0$ and at $ Time Steps = 60\times10^3 (a), 70\times10^3
(b), 80\times10^3 (c), 84\times10^3 (d), 86\times10^3 (e), 90\times10^3 (f),
94\times10^3 (g), 96\times10^3 (h), 10^5 (i)$
}
\label{Nucl}
\end{figure*}

Clusters within the coexistence state appear through a nucleation and growth mechanism,
as modified by the anisotropy induced by the presence of the drive. Fig.~\ref{Nucl} shows that for a particular disorder realization at a force $F_x=98$, a 
nucleus of square region appears  which grows with time.
In all the disorder realizations we studied, the square nucleus  first formed in the hexagonal phase is anisotropic
and elongated along the transverse direction (Fig.~\ref{Nucl}), a consequence of the fact that
the drive introduces a preferred direction into the system.

\begin{figure*}
\begin{center}
\includegraphics[width=15.0cm]{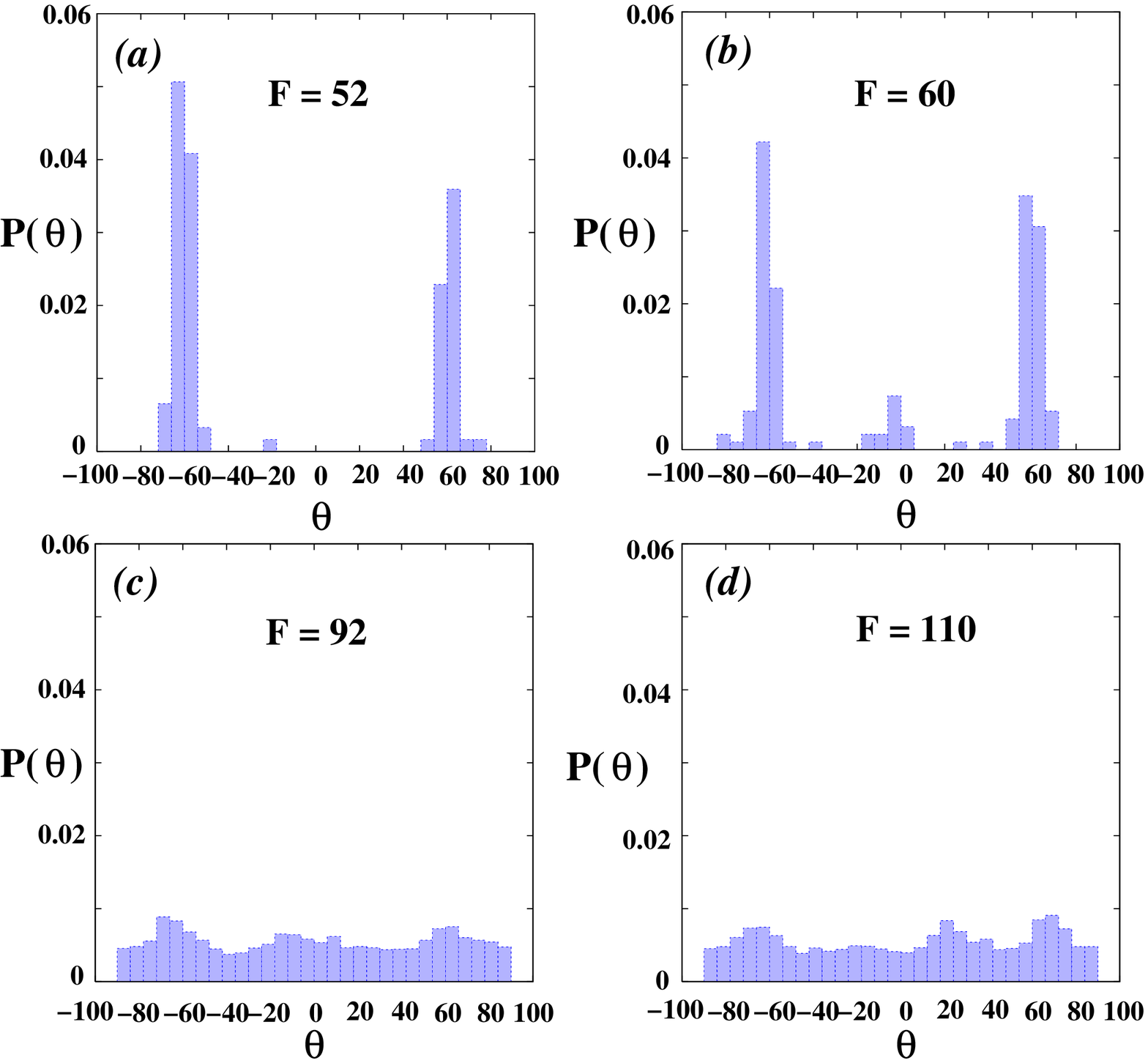}
\end{center}
\caption{[Color Online:] 
The distribution of the angle made by the major principal axis (the eigenvector corresponding to the eigenvalue $\lambda_>$)
of the square islands within a triangular background, computed for different values of the driving force $F$.
The force values are (a) F = 52, (b) F = 60, (c) F = 92 and (d) F = 110,
with $v_3 = 6$.}
\label{quenchNucl}
\end{figure*}

Square clusters in the coexistence regime tend to be less anisotropic compared to the elongated 
string-like defects of square within a triangular background obtained at smaller force values. This can be seen from 
Fig.~\ref{quenchNucl} which plots the distribution of angles made, with respect to the drive direction,  by the
principal direction of  the square nucleus, corresponding to the larger eigenvalue of the moment of inertia
tensor. For small $F$, the distribution peaks around
$\theta = \pm 60^\circ$, indicating that the square nucleus forms preferentially at a $60^\circ$ angle
with respect to the drive. For forces within the coexistence regime, however, the distribution of this
angle is smooth and has no sharp peaks, concomitant with the shape of the nucleus becoming more isotropic.

\subsubsection{Noise in the Coexistence Regime}
The  interconversion  between square and triangular regions leads to
complex spatio-temporal behavior. To demonstrate this,  we  compute the power 
spectrum of fluctuations of the particle current.
We first obtain the statistics of the number
of particles crossing an imaginary line parallel to the
transverse ($y$) direction per unit time, per unit length of
the line for a particular force ($F_x$) value. The power spectrum
$S_{flux}(f)$ is the Fourier transform of
the auto-correlation function of this time series of particle flux,
averaged over different choices of the position of the imaginary
line and over different choices of the time slots of observation.
Considerable statistics were taken to ensure that all quantities
were well averaged.  About $5$ different choices of the position of the transverse 
imaginary line were taken, in addition to $20$ different choices of time
slots. Averaging is performed  over a large span of simulation time steps ($\sim 10^9$).

\begin{figure}
\begin{center}
\includegraphics[width=8.0cm]{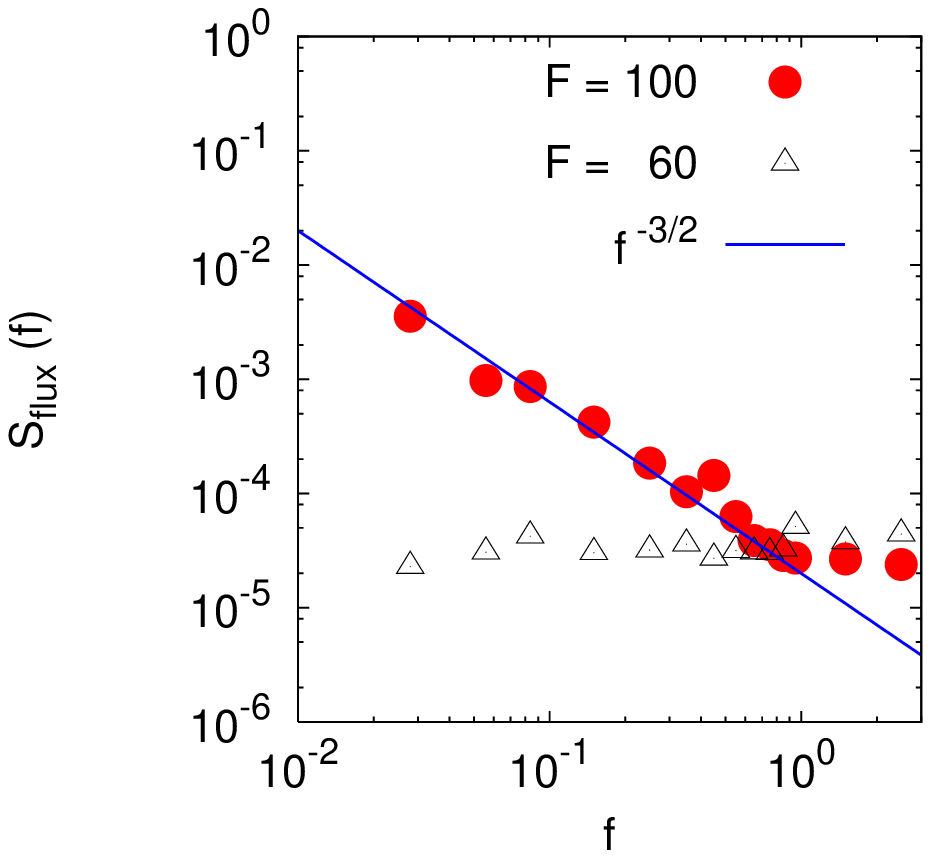}
\end{center}
\caption{[Color Online:] The power spectrum $S_{flux}(f)$
of current fluctuations in the moving triangle phase (open triangles $F_x=60$) 
and in the coexistence region (red circles $F_x=100$), logarithmically
binned and plotted for a range of frequencies below the washboard 
frequency.  Current fluctuations in the coexistence region
are {\it enhanced}, also showing a $1/f^{3/2}$ decay in the low frequency
range. 
}
\label{autocor}
\end{figure}
\begin{figure*}
\begin{center}
\includegraphics[width=14.0cm]{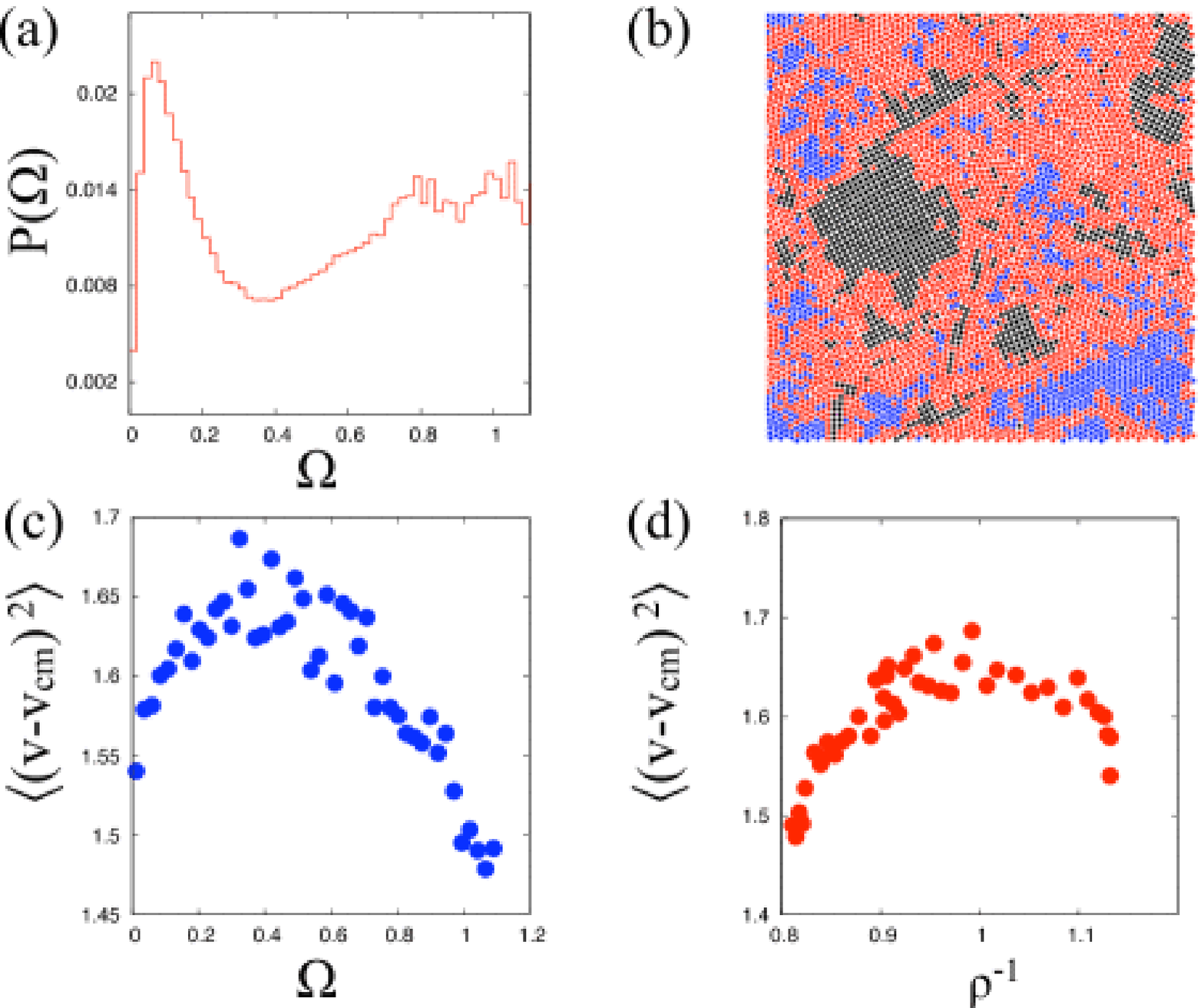}
\end{center}
\caption{[Color Online:] (a) Probability distribution of the bond angle order parameter $\Omega$ (see text) in the coexistence regime. The peak at low value of $\Omega$ corresponds to the square phase. The series of peaks at high $\Omega$ values all correspond to various combinations of bond angles in the (disordered) triangular phase. The interfacial region has $\Omega$ corresponding to the first dip in the curve.  (b) Illustration of the spatial distribution of $\Omega$ values as stated in (a) $\Omega < 0.2$ (square) are black, those for $1.3 >\Omega > 0.2$ 
are red  (interface) and those for $\Omega > 1.3$ are blue (triangle).  (c) The fluctuation of the local velocity about the centre of mass value $<(v - v_{cm})^2>$ plotted against $\Omega$. It is clear that the fluctuations are largest for the particles in the interfacial region. (d) The fluctuation of the local velocity as a function of the local volume per particle. }
\label{omegas}
\end{figure*}

Our results are summarized  in Fig.~\ref{autocor} for two values
of $F_x$. For $F_x = 60$, when the system is in the moving triangular
phase, $S_{flux}(f)$ is, to a large extent,  featureless and flat,
except for large $f$ where effects due to the streaming of the entire
system become important. In contrast, for a high force $F_x = 100$, when the system
is within the coexistence phase, we obtain a $1/f^\alpha$ regime, with $\alpha \simeq 1.5$ in
$S_{flux}(f)$ over about a decade.  In addition the particle current fluctuations remarkably
{\it enhanced } by  3-4 orders of magnitude in the coexistence phase
as compared to the triangular regime.

\subsubsection{Origins of Noise in the Coexistence Regime}

To investigate the origins of the noise in the coexistence regime, we define the local quantity
 \begin{equation}
 \Omega_i = \sum_{j,k}\sin^2(4 \theta_{ijk}),
 \end{equation}
where the summation is over a defined
region ($r < r_0 = 1.2 \sigma$) surrounding a particle i. Here $\theta_{ijk}$ is the bond angle
between particle $i$ and $j$ and $k$. We choose  particles $j$ and $k$ such that they are all within a 
specified cutoff radial distance from particle i.

Fig.~\ref{omegas}(a)  displays the probability distribution of $\Omega \equiv  \Omega_i$ for a state in the coexistence region. There are two prominent peaks. 
The peak for small values of $\Omega$ corresponds to the square lattice ($\Omega = 0$ for the ideal square crystal), whereas
 the peak at higher values of $\Omega$ corresponds to the triangular structure. Intermediate values of $\Omega$ are obtained for particles at the 
 interface between square and triangle. 
 Fig.~\ref{omegas}(b) shows a  snapshot of the system in the coexistence region with particles color coded according to their value of $\Omega$.  
 
 We next examine  the fluctuations of the velocity about the average value $\langle (v - v_{cm})^2 \rangle$
 as a function of the local coordination. In Fig.~\ref{omegas}(c) this is plotted as a function of $\Omega$. We see  that velocity fluctuations are 
 larger in the interfacial region. Further, Fig.~\ref{omegas}(d) shows a plot of the local volume per particle as a function of the velocity 
 fluctuations,  showing that  the relatively lower density of the interfacial region (due to the presence of a large concentration of defects) causes it to fluctuate more than the rest of the solid.  The rapid fluctuations of the interface also results in rapid interconversion between square and triangular coordinated particles and leads also to enhanced fluctuations in the coordination number.    Since the driven solid is elastically constrained to move as a whole in the direction of the drive without gaps or cracks, this sets up strong correlations among the particle trajectories. Such correlations are ignored in the theory of the shaking temperature of Koshelev and Vinokur.
\begin{figure*}
\begin{center}
\includegraphics[width=15.0cm]{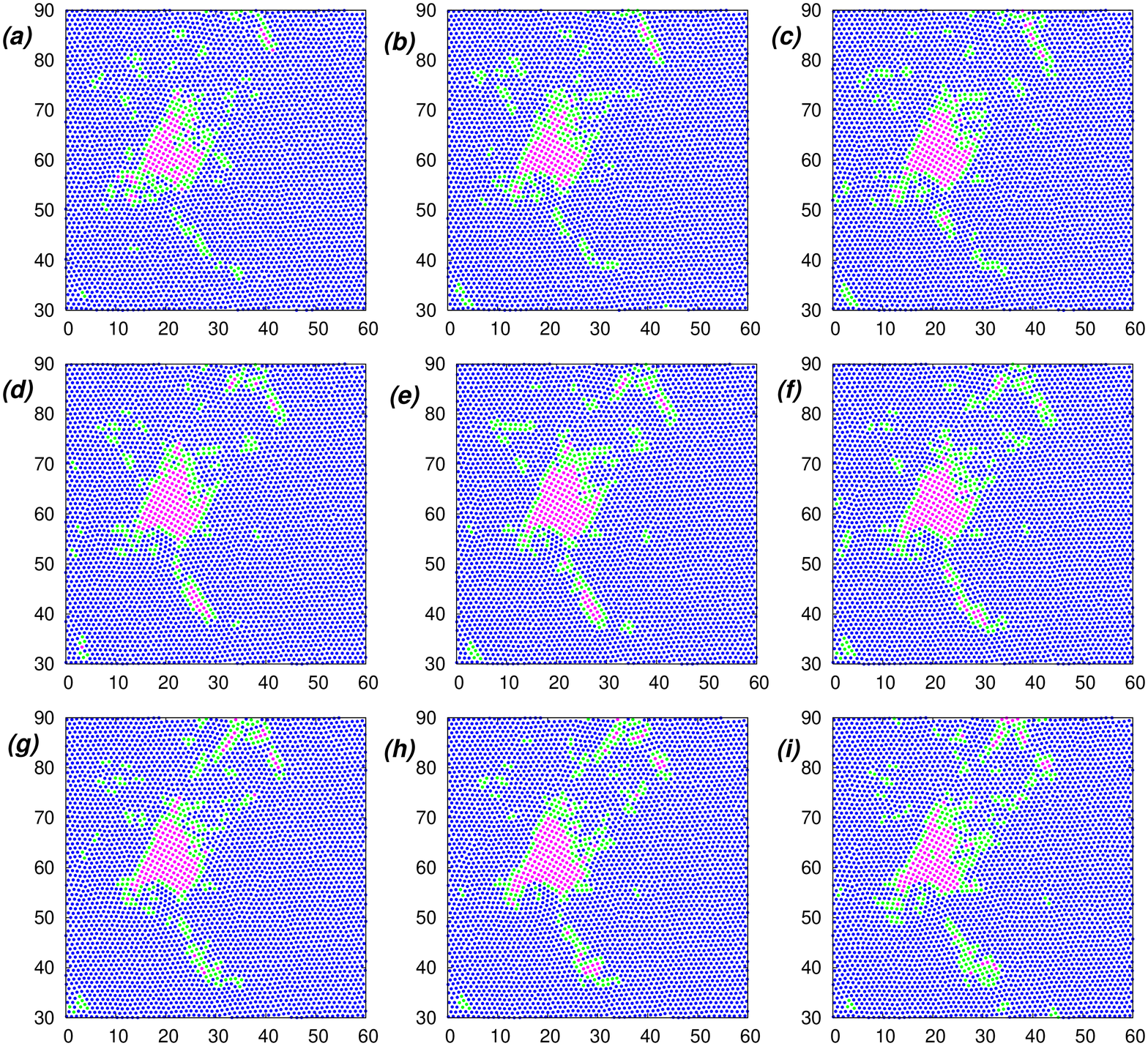}
\end{center}
\caption{[Color Online:] Configuration snapshots arising from a quench to zero force 
of a randomly chosen configuration within the
coexistence regime, at time steps (a) 0, (b) 4$\times10^3$,
(c) $10^4$, (d) 5 $\times  10^4$, (e) $10^5$, (f) $2 \times 10^5$, (g) $3 \times 10^5$, (h) $5 \times 10^5$,
and (i) $9 \times 10^5$, with $v_3 = 6.0$. The system size is N = 10000.}
\label{quenchNucl2}
\end{figure*}
\section{Dynamical Coexistence and Peak Effect Anomalies}

In this section, we turn to a possible application of this model, in the context of
experiments on transport anomalies associated with 
peak effect phenomena in the superconducting mixed phase. To recapitulate, the peak effect refers to the 
sharp increase in the critical current $j_c$
in the mixed phase of a  disordered type-II superconductor close to H$_{c2}(T)$. 
This critical current is  thus a non-monotonic function of
$T$ (or $H$, depending on which is varied experimentally), since
it  decreases steadily from its low temperature value till the onset temperature $T_o$ 
of the peak effect, from whence it increases sharply over a small
temperature range to its maximum value, obtained at temperature $T_m$. As $T$
is raised still further, the critical current collapses again, to reach  zero
in the normal state. The temperature interval  [$T_0,T_m$] in which the critical current rises 
anomalously is the peak effect regime. This regime is dynamically
anomalous, displaying: (i) large current noise amplification at low frequency, (ii) a $1/f$ spectrum
of current fluctuations, which is very non-Gaussian, (iii) a ``fingerprint''
effect in which apparently random spikes in the differential resistivity as a function
of drive are retraced as the drive is decreased, (iv) a history-dependent dynamic response (v) a memory of
direction, amplitude and frequency of applied currents, (vi) a strong suppression of ac response by
a dc bias  as well as a variety of other behaviour\cite{shobo1, shobo2, shobo3, shobo4, shobo5, shobo6,ghosh, ravi1,satyajit1, satyajit2, satyajit4,andrei1,andrei2}.
A very large number of experiments probing such anomalous
behavior, including all those referenced above, are transport-based, thus
serving as 
probes of  the dynamics of vortices within this narrow region of parameter space.

Interpretations of these phenomena are largely phenomenological.
One particularly influential proposal considers
an underlying order-disorder transition ``contaminated'' by
sample surfaces or ``edges''. Such surfaces, with associated surface barriers for vortex entry,
provide an entry point for vortices driven into flow\cite{beidenkopf}. 
The surface should provide an intrinsically  more disordered environment for
vortices than the bulk, particularly in fairly pure samples where $j_c$ is low. Thus, vortices might
be expected to enter through the boundaries in a highly disordered state, only to anneal 
in the nearly pure bulk, when a current is applied across the sample.  
This spatial separation of disordered and ordered states and the 
slow annealing of  one into the other is argued to be the central feature
underlying the anomalous behavior seen in the peak effect regime. 
Magneto-optic imaging via Hall bar arrays support the surface contamination scenario.
However, such methods do not access the dynamics of annealing and phase transformations
directly. Much recent work appears consistent with a bulk coexistence of disordered and
ordered phase\cite{pasquini2008}, while decoration experiments see a ``multi-domain''
structure in the peak effect regime\cite{menghini2002,fasano2002}, as proposed in 
Refs.~\cite{menon1,menon2,menon3} and accessed indirectly in Refs.~\cite{first04,thirdmom}.
For related simulations, see Ref.~\cite{moretti05}.

The  edge contamination scenario implicitly assumes that the underlying order-disorder
transition is unaffected by the drive, serving only to provide a background to the annealing
process.  However,  in a generic driven system, the possibility that the drive has 
more non-trivial effects must be expected. In particular, the drive may alter the very nature of the
driven bulk, stabilizing dynamical states that are truly non-equilibrium in character, as
illustrated in the simulations discussed above.

A  beautiful recent experiment  (Ref.~\cite{shobonature}) performs  a variant of scanning probe 
microscopy, using a mounted local Hall probe. The probe
is  sensitive to variations in the local magnetic induction averaged across a mesoscopic
scale of around a micron. The system is
tuned across the peak effect regime and then perturbed weakly through
a low-amplitude ac field applied from below the sample. The Hall probe,
placed above the sample and linked, though lock-in techniques to the
frequency of the ac perturbation, records a local  susceptibility,
indicative of pinning response, as a function of space and integrated over the 
thickness of the sample. The spatial resolution is limited by the size of
the Hall probe, typically of the order of a micron or so in size.

As parameters are varied across the peak effect,  these experiments see
a remarkable coexistence  between a strong pinning regime and a weak 
pinning regime.  A complex interface is seen between these coexisting states with a
dynamics which is exquisitely sensitive to the field and the disorder,
Such coexistence is also a feature of other phenomenological approaches to this
problem, which address transport measurements.
Locally more disordered regions of the sample 
appear to nucleate more stable  regions of strong pinning whereas small variations of
the applied field  cause large changes in the inhomogeneous 
pinning pattern. However, the complex geometry of the
coexisting regimes appears largely stable if the temperature and field are fixed, suggesting
that thermal fluctuations are not dominant. While
vortices entering from the sample boundaries do appear to contribute 
to this dynamics in no small measure, there is  significant evidence for non-trivial dynamics 
in the bulk, with regions of strongly pinned phase being nucleated far from any boundary. Thus,
these experiments point to a more active role for the bulk than envisaged in the 
boundary injection scenario. In this context, the authors of Ref.~\cite{shobonature,vortcap} have 
specifically argued that the  complex topology of the two-phase interface should be
largely  responsible for the  history dependence seen in the experiments.

How are these remarkable observations related to the model we
study here? We suggest that the link is the {\em emergence of a self-organized, 
disorder-stabilized, dynamically sustained drive-induced coexistence phase} seen
both in the experiments and in simulations  of our model 
system. The  similarities between the two are striking:
First, the coexistence itself. Both the experiments and our simulations here provide 
incontrovertible evidence for dynamical states in driven disordered systems
which resemble phase coexistence at equilibrum phase transitions, with the
added complication of spatial inhomogeneities due to quenched disorder\cite{valenzuela,pasquini1,pasquini2008,jang2009}. Reasoning
from the experiments, the complex interconversion of one phase into the other and
the spatially inhomogeneous character of the dynamics is the hallmark of vortex
dynamics within the peak effect regime\cite{shobonature}. This is precisely the situation which obtains in
the simulations. As pointed out in the previous section, what is unusual about the
coexistence regime is that the strong disorder-induced fluctuations seen and manifest in all the dynamic
properties we measure are obtained {\em above} the depinning transition, surviving
even at large values of the applied force. 

Second,  the coexistence seen in the experiments is very disorder-sensitive\cite{shobonature,jaiswal}. In experiments,
this is manifest in terms of the complex structure of differentially pinned regions
in the sample, presumably reflecting a non-trivial pinning landscape. We see similar dynamical 
behavior within the coexistence regime, with the structure at fixed drive influenced by
the underlying microscopic disorder and sensitive to even marginal changes in the pinning. Thus, the
nature of inhomogeneities connected to dynamical phase coexistence in this model 
as well as in the experiments appears to be  dictated primarily by the underlying disorder. 

Third, the slow dynamics and long relaxation times seen in the experiments, reflecting the
complex dynamics of the interface separating the regimes of different pinning strength,
is seen in our simulations as well\cite{shobophysica,thakur}. In our simulations, if the drive is switched off and the system 
allowed to anneal, the nuclei of one phase within the other  live anomalously long,
suggesting that the dynamics has stabilized long-lived metastable states, precisely
as seen in the experiments. In Fig.~\ref{quenchNucl2} we show the
dynamics of a square nucleus within a triangular background, extracted from a typical
configuration within the coexistence regime,
when quenched to zero drive. Far from vanishing over a short time scale, 
the nucleus appears to be remarkably stable out to the longest time scales accessed in our simulations, 
providing evidence that the metastability exhibited in the coexistence regime survives even
after the drive is turned off.  These are reflected in the memory experiments, in which the 
removal of the driving transport current essentially appears to freeze the system
into a metastable state from which it only recovers upon a reapplication of the
drive.

Fourth and finally, the noise spectra within the coexistence regime, including the anomalously
large noise and the power-law fall-off with a $1/f^\alpha$ spectrum is a prominent 
feature of the peak effect regime. Experiments see a power-law falloff,  with $1  \leq \alpha \leq 2$, 
as well as a substantial enhancement in noise power of a few orders of magnitude\cite{shobo4,shobo5}. 
The substantial non-Gaussian features in the noise, as obtained in Ref.~\cite{shobo5}, indicate
that a small number of fluctuators contribute; it is tempting to assign these to a few large domains
which fluctuate collectively within the coexistence regime, as suggested by the simulations. 

The edge contamination scenario {\em  assumes} that the drive in the bulk anneals the disordered
vortices injected across the boundaries into the smoothly flowing state. In contrast, we show that 
in the vicinity of an underlying transition, the bulk flowing state is generically inhomogeneous and 
dynamically non-trivial, thereby questioning this basic assumption. We believe that the edge contamination scenario
should apply, more immediately, close to but away from the peak effect regime, where the drive in the bulk
acts solely to smoothen flow, precisely as seen in a recent experiment\cite{mohan}. 

We stress that the system we simulate is far from a literal translation of the vortex system. 
It lacks the  $K_0(r/\lambda)$, (with $\lambda$ the
penetration depth), interactions of vortex lines, which can become fairly long-range
if $\lambda$ is large. In our simulations we drive transitions between the two
different pure system phases by varying a parameter $v_3$, whereas these transitions
are density or temperature driven in the classic scenario of the peak effect in vortex systems. (The underlying 
order-disorder transition in the vortex system is likely related to a melting-like transition, at least for small $H$
and for temperatures close to $T_c$ and not a structural transition between two crystalline phases as in our model\cite{menon3}.)
The experiments are in three dimensions, whereas our simulations are two-dimensional. Our system shows
very little variation in the depinning force close to the transition, and thus no apparent peak effect. (It  is known,
however, that factoring in the temperature dependence of the coherence length $\xi$ and $\lambda$
is essential to obtain a peak effect in two-dimensional simulations\cite{chandran}.)

For these, as well as other reasons,  our proposal for the origin of peak effect anomalies should be thought of 
as indicating a generic scenario within which much of the observed physics finds a common explanation. 
Clearly more work, including simulations of the three-dimensional case capable of describing
vortex entanglement, is required to further illuminate  the relationship conjectured here.

\section{Summary and Conclusion}

This paper has studied the complex dynamical behavior of a 
two dimensional driven disordered solid which undergoes a square to 
triangular  structural transition as a single parameter $v_3$ is tuned. 
Our interest in this model has several origins: The problem of
dynamical probes of an underlying static phase transition in a weakly
disordered system is interesting in itself. The possibility that
such behavior might underlie a classic and  ill-understood problem
in the literature on type-II superconductivity, the problem of the origin
of peak-effect anomalies, provides further motivation for this study. 

Specifically, using the initial study of Ref.~\cite{ankush}  and this paper, we have demonstrated the following: 
First, the existence of a {\em complex phase diagram containing
a large number of states induced purely by the drive}, such as the anisotropic hexatic
state and the coexistence state. Second,   
the demonstration of {\em dynamically unusual behaviour in the coexistence region}, as measured 
through several dynamical quantities, including the observation of $1/f^\alpha$ correlations
in the current noise, the existence of highly metastable states  and 
the observation of a very substantial noise enhancement.
Third,  the {\em breakdown of single-particle, effective temperature 
descriptions in the coexistence regime}, where measures of a flow and disorder induced temperature
indicate  substantially non-monotonic variation within the coexistence regime. Fourth,  the demonstration that 
the {\em putative driven hexatic
glass phase cannot have algebraically decaying orientational correlations}, since correlations 
along the drive direction are always long-ranged while those along the transverse direction are 
generically short-ranged. Fifth, the intuitively unexpected {\em  expansion of the plastic flow regime at large $v_3$ and its
subsequent collapse}. Finally, we have suggested  a {\em possible connection
to the classic problem of the origin of peak effect anomalies}, conjecturing that the explanation might be found
in the very nature of the unusual dynamical states obtained when a system close to a first-order
(structural or melting) phase transition is driven across a quenched-disordered background. 

There are several realizations
of non-equilbrium states of complex fluids, such as sheared lamellar phases , worm-like micelles, and
driven colloidal suspensions, which exhibit remarkably non-trivial  behaviour as a
consequence of dynamical phase transitions\cite{rheochaos1,rheochaos2,rheochaos3}. Possible relations to
equilibrium descriptions of systems at phase coexistence have also been outlined\cite{rheochaos4}. Many features,
including profoundly non-linear response, complex spatio-temporal behaviour and
large noise signals appear to be common to these systems when driven away from equilibrium\cite{rheochaos1}. 
What singles out the category of systems we study here is the additional complication of quenched disorder.
Clearly these are suggestive  links at the intersection of these various fields. Further exploration of these connections
would appear to be fruitful.

\begin{acknowledgements}
The authors thank S. Bhattacharya, D. Dhar and C. Dasgupta
for discussions. In addition, GIM is grateful to  S. Bhattacharya for many interactions
over the years concerning the phenomenology of the peak effect. The work of all three authors
was supported in large part by the DST (India). AS thanks S. N. Bose National Centre for Basic Sciences
for computational facilities and financial support, and gratefully acknowledges ongoing  support  from 
SFB TR6 (DFG).
\end{acknowledgements}

\end{document}